\documentclass{aa}
\usepackage{graphicx}
\usepackage{dirtytalk}
\usepackage{txfonts}
\usepackage{hyperref}
\usepackage{xcolor}
\hypersetup{
colorlinks,
linkcolor={red!80!black},
citecolor={blue!80!black},
urlcolor={blue!80!black}
}

\newenvironment{myfont}{\fontfamily{ppl}\selectfont}{\par}
\DeclareTextFontCommand{\textmyfont}{\myfont}

\newcommand{\code}[1]{\texttt{#1}}

\def\nifs{\iso{56}Ni}

\def\cofs{\iso{56}Co}

\def\fefs{\iso{56}Fe}

\def\cm3{cm$^{-3}$}
\def\kms{\mbox{km~s$^{-1}$}}
\def\msunyr{$M_{\odot}$\,yr$^{-1}$}

\def\mdot{$\dot{\rm M}$}

\def\msun{$M_{\odot}$}

\def\vinf{$V_\infty$}

\def\one{\ts {\,\sc i}}
\def\two{\ts {\,\sc ii}}

\def\beq{\begin{equation}}
\def\eeq{\end{equation}}

\def\lesssim{\mathrel{\hbox{\rlap{\hbox{\lower4pt\hbox{$\sim$}}}\hbox{$<$}}}}
\def\gtrsim{\mathrel{\hbox{\rlap{\hbox{\lower4pt\hbox{$\sim$}}}\hbox{$>$}}}}

\def\one{{\,\sc i}}
\def\two{{\,\sc ii}}

\def\v1d{{\code{V1D}}}

\def\cmfgen{{\code{CMFGEN}}}
\def\heracles{{\code{HERACLES}}}

\def\mej{$M_{\rm ej}$}

\def\ergs{erg\,s$^{-1}$}

\def\caiidoub{[Ca\two]\,$\lambda\lambda$\,$7291,\,7323$}

\def\oidoub{[O\one]\,$\lambda\lambda$\,$6300,\,6364$}

\newcommand{\iso}[2]{\ensuremath{^{#1}\rm{#2}}}

\begin{document}

\title{Helium stars exploding in circumstellar material and the origin of Type Ibn supernovae}
\titlerunning{The nature of SNe Ibn}

\author{
Luc Dessart\inst{\ref{inst1}}
 \and
  D. John Hillier\inst{\ref{inst2}}
 \and
Hanindyo Kuncarayakti\inst{\ref{inst3},\ref{inst4}}
 }

\institute{
    Institut d'Astrophysique de Paris, CNRS-Sorbonne Universit\'e, 98 bis boulevard Arago, F-75014 Paris, France.\label{inst1}
\and
   Department of Physics and Astronomy \& Pittsburgh Particle Physics, Astrophysics, and Cosmology Center (PITT PACC),  \hfill \\ University of Pittsburgh, 3941 O'Hara Street, Pittsburgh, PA 15260, USA.\label{inst2}
\and
    Tuorla Observatory, Department of Physics and Astronomy, FI-20014 University of Turku, Finland.\label{inst3}
 \and
     Finnish Centre for Astronomy with ESO (FINCA), FI-20014 University of Turku, Finland.\label{inst4}
  }

   \date{}

   \abstract{
   Type Ibn supernovae (SNe) are a mysterious class of transients whose spectra exhibit persistently narrow He\one\ lines, and whose bolometric light curves are typically fast evolving  and overluminous at peak relative to standard Type Ibc SNe. We explore the interaction scenario of such Type Ibn SNe by performing radiation-hydrodynamics and radiative-transfer calculations. We find that standard-energy helium-star explosions within dense wind-like circumstellar material (CSM) can reach on day timescales a peak luminosity of a few $10^{44}$\,\ergs, reminiscent of exceptional events like AT\,2018cow. Similar interactions but with weaker winds can lead to Type Ibc SNe with double-peak light curves and peak luminosities in the range $\sim$\,$10^{42.2}$ to $\sim$\,10$^{43}$\,\ergs. In contrast, the narrow spectral lines and modest peak luminosities of most Type Ibn SNe are suggestive of a low-energy explosion in an initially $\lesssim$\,5\,\msun\ helium star, most likely arising from interacting binaries, and colliding with a massive helium-rich, probably ejecta-like, CSM at $\sim$\,10$^{15}$\,cm. Nonlocal thermodynamic equilibrium radiative-transfer simulations of a slow-moving dense shell born out and powered by the interaction compare favorably to Type Ibn SNe like 2006jc, 2011hw, or 2018bcc at late times and suggest a composition made of about 50\% helium, a solar metallicity, and a total ejecta/CSM mass of 1--2\,\msun. A lower fractional helium abundance leads to weak or absent He\one\ lines and thus excludes more massive configurations for observed Type Ibn SNe. Further, the dominance of Fe\two\ emission below 5500\,\AA\ seen in Type Ibn SNe at late times is not predicted at low metallicity. Hence, despite their promising properties, Type Ibn SNe from pulsational-pair instability in very massive stars, which requires low metallicity, have probably not yet been observed.
}
\keywords{Stars: evolution -- radiative transfer -- supernovae: general -- supernova: individual: SN\,2006jc, SN\,2011hw, SN\,2018bbc, AT\,2018cow}

   \maketitle

\section{Introduction}

Supernova (SN) 1999cq was the first reported occurrence exhibiting typical Type Ic spectral properties superimposed with the presence of narrow He\,\one\ lines, suggestive of the interaction of an hydrogen-free ejecta with helium-rich hydrogen-free circumstellar material \citep[hereafter CSM;][]{matheson_99cq_00}. Type Ibn SNe discovered since then reveal a diversity in photometric and spectral properties. This includes events such as  SN\,2006jc \citep{foley_06jc_07,pasto_06jc_07}, SN\,2005la \citep{pasto_05la_ibn_08}, SNe 2010al and 2011hw \citep{smith_11hw_12,pasto_ibn_trans_15}, OGLE-2012-SN-006 \citep{pasto_ibn_slow_15}, SN\,2014av \citep{pasto_14av_16}, ASASSN-15ed \citep{pasto_15ed_ibn_15}, or SN\,2015U \citep{shivvers_ibn_16}. The peak luminosity of Type Ibn SNe can reach values as high as in Type IIn SNe, covering the range 10$^{42}$--10$^{43.5}$\,\ergs\ (corresponding to peak $R$-band magnitudes in the range $-17$ to $-20$\,mag; \citealt{pasto_ibn_08}, \citealt{hosseinzadeh_ibn_17}). The rise time to maximum is usually short (much shorter than in Type IIn SNe of similar maximum luminosity), ranging from a few days to $\sim$\,15\,d. A single light curve maximum is normally witnessed but some events exhibit a double-peak light curve (e.g., SN\,2011hw; \citealt{pasto_ibn_trans_15}). The post-maximum decline rates are usually high although in some rare cases like OGLE-2012-SN-006 the luminosity remains substantial out to late times \citep{pasto_ibn_slow_15}. Type Ibn SNe have blue spectra prior to and at maximum light, with lines of He\,\one\ that greatly vary in shape (pure emission line or P-Cygni profile) and width between objects and between epochs. SN\,2006jc  exhibited He\,\one\ line widths suggestive of an expansion speed of about 2000\,\kms. SN\,2010al and ASSAS-15ed showed He\,\one\ lines that were narrow up to maximum (full-width-at-half-maximum, hereafter FWHM, of $\sim$\,1000\,\kms),  and broadened with time (FWHM $\sim$\,5000\,\kms). Although not the norm, H\,\one\  lines can be present (e.g., SN\,2005la; \citealt{pasto_05la_ibn_08}). Narrow lines with broad wings resulting from non-coherent scattering with free electrons are sometimes seen in Type Ibn SNe (e.g., SN\,2010al; \citealt{pasto_ibn_trans_15}), although this seems much rarer than in Type IIn SNe (some representative Type Ibn SN spectra are reproduced in Sect.~\ref{sect_obs} of this paper).

The general consensus on the origin of Type Ibn SNe is a Wolf-Rayet star exploding within a helium-rich CSM \citep{matheson_99cq_00, pasto_06jc_07,hosseinzadeh_ibn_17}. Lacking a good theory for the pre-SN mass loss, it is not clear what progenitors produce these Type Ibn SNe. Theoretical models of high mass stars suggest that pulsational-pair instability in very massive stars at low metallicity are ideal progenitors for Type Ibn SNe \citep{yoshida_ppisn_16,woosley_ppsn_17}. However, observations of the explosion site and its environment suggest that in some cases, lower-mass massive stars are favored, for example for SN\,2006jc or 2014C \citep{maund_06jc_16,sun_06jc_20,sun_14C_20}.   In some cases, the association of Type Ibn SNe with massive stars is even lacking \citep{hosseinzadeh_ibn_19}. Support for the interaction scenario is in part given by the UV and X-ray detections, for example for SN\,2006jc \citep{immler_06jc_08}. Emission from a slow-moving dense shell yields a satisfactory explanation for the spectral characteristics of Type Ibn SNe \citep{chugai_06jc_09}. Numerous studies promote the interaction of an ejecta of 10$^{51}$ to 10$^{52}$\,erg with CSM (as for SN\,2006jc; \citealt{tominaga_06jc_08}; \citealt{chugai_06jc_09}), but this scenario conflicts with the lack of broad spectral lines and the relatively modest peak luminosities of Type Ibn SNe.

To clarify the origin of Type Ibn SNe and investigate the physical conditions producing their  spectral properties, we conduct a number of numerical simulations for the radiation hydrodynamics of ejecta/CSM interaction and for the nonlocal thermodynamic equilibrium (non-LTE) radiative-transfer in a hydrogen-free helium-rich dense shell powered by interaction. In the next section, we discuss the representative mass of helium-rich material present in massive stars at the end of their lives. This sets the stage for the potential ejecta/CSM configurations relevant for Type Ibn SNe. Section~\ref{sect_rhd} presents results for a set of radiation-hydrodynamics simulations of the interaction between an ejecta and CSM either produced in a non-terminal explosion or in a super-wind phase. Section~\ref{sect_rt} presents non-LTE radiative transfer calculations for a slowly moving dense shell subject to a prescribed power. After presenting the numerical setup (Sect.~\ref{sect_rt_setup}), we discuss the basic properties for a representative configuration (Sect.~\ref{sect_he4p0}), the dependency of radiative properties on the cold-dense shell (CDS) mass and composition (Sect.~\ref{sect_mass}), and the impact of the iron abundance  (Sect.~\ref{sect_iron}). We present a few spectral comparisons to observed Type Ibn SNe in Sect.~\ref{sect_obs} and conclude in Sect.~\ref{sect_conc}.

\section{How much helium in massive stars at core collapse}
\label{sect_he}

The dominance and persistence of He\one\ lines is a fundamental property of Type Ibn SNe. The absence of H\one\ lines, as observed in Type IIn SNe, shows that the helium is not from the hydrogen-rich mixture encountered for example in the envelope of red-supergiant stars. Type Ibn SNe must arise from explosions or interactions in which helium dominates the composition. Since the mass of helium will be a fundamental parameter affecting the spectra of Type Ibn SNe, we quantified how much helium, free of hydrogen, is present in massive stars at the time of core collapse.\footnote{Some stars can contain a lot of helium but much earlier in their evolution. For example, helium stars, formed after prompt removal of the hydrogen-rich envelope by Roche lobe overflow at the onset of core helium burning, are nearly pure helium (see, for example, the initial conditions for such helium stars in \citealt{woosley_he_19}). However, an eruption or an explosion is not expected in such objects at this phase of evolution.}

\begin{figure}
\centering
\includegraphics[width=\hsize]{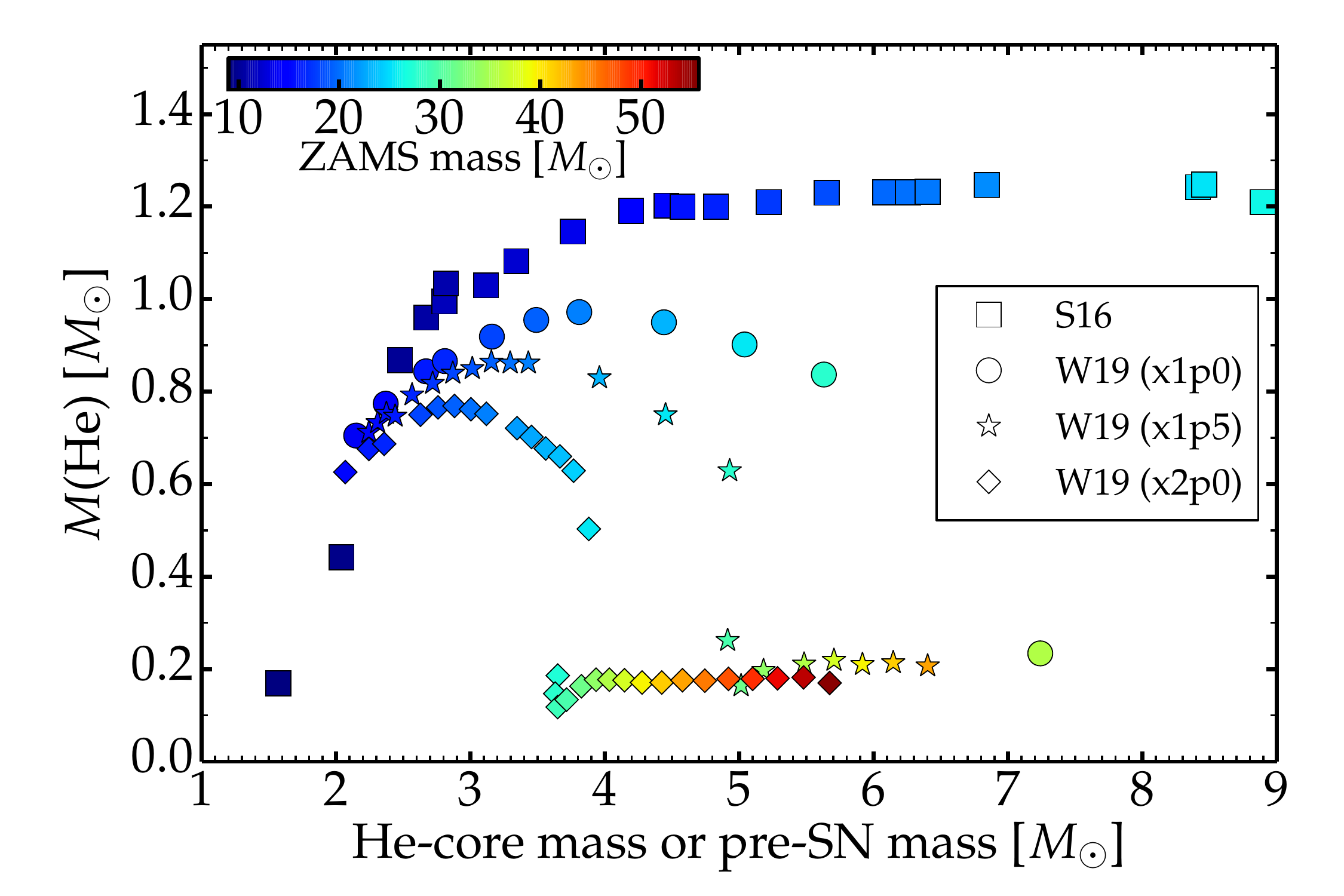}
\caption{Illustration of the helium content of massive star progenitors at core collapse and evolved either in isolation or in an interacting binary system. For single-star progenitors, we use the models of \citet{sukhbold_ccsn_16}, labelled S16, and show the combined masses of the He/C and the He/N shells in the progenitor versus the helium-core mass, both taken at the time of core collapse. For the binary star progenitors, we use the helium-star models of \citet{woosley_he_19}, labelled W19 (we include models evolved with three different mass loss rates), and show the total helium yields versus pre-SN mass. The color coding indicates the ZAMS mass for each model. [See Sect.~\ref{sect_he} for discussion.]
\label{fig_he}
}
\end{figure}

Figure~\ref{fig_he} is a compilation of results from single- and binary-star evolution calculations. The single-star models are from \citet{sukhbold_ccsn_16} and cover main-sequence masses between 9.0 and 26.5\,\msun. These models reach core collapse with a massive residual hydrogen-rich envelope, an helium core unaffected by mass loss, and a combined He/C and He/N shell mass typically between 0.8 and 1.2\,\msun\ for 10--26.5\,\msun\ progenitor masses, and down to 0.2-0.4\,\msun\ for progenitors of 9.0 and 9.5\,\msun. With the exception of the lightest massive stars undergoing core collapse, there is a narrow range of helium-shell masses, which results from the fact that higher mass massive stars have an ever growing CO core so that the He/C and He/N shell masses are similar in a 10\,\msun\ and a 26.5\,\msun\ star. The binary-star models in Fig.~\ref{fig_he} are from \citet{woosley_he_19} and cover initial helium-star masses between 2.6 and 22\,\msun\ (corresponding to main-sequence masses between $\sim$\,14 and $\sim$\,56\,\msun). Three model subsets are shown that assume a nominal mass loss rate (x1p0), a 50\% (x1p5) and a 100\% (x2p0) enhanced mass loss rate during the helium-star evolution. Because of the effect of mass loss on the helium core and the lack of an hydrogen-burning shell adding helium to the helium core (as in single stars), the final helium yield is lower that in the single star set, and between 0.7--1.0\,\msun\ for models he2p6 to he5p0 and decreasing for higher initial helium-star masses.  This reduction in helium mass is greater for higher adopted mass loss rates.

Figure~\ref{fig_he} shows that no model, however massive, ejects more than about 1.2\,\msun\ of helium (unmixed with hydrogen), and that the models with the largest fractional helium abundance are lower mass helium stars in binaries (initial helium-star mass between 2.6 and 5.0\,\msun). This is a strong hint that the progenitors of Type Ibn SNe might be found in lower-mass massive stars evolving in interacting binaries.

At the high mass end, pulsational-pair instability models are predicted to eject shells that contain helium, but in these ejecta helium represents only a small fraction of the total mass, typically less than $\sim$\,20\,\% \citep{yoshida_ppisn_16}. This composition may not be suitable to explain the dominance and persistence of He\one\ lines observed in Type Ibn SNe -- we will return to this point in Sect.~\ref{sect_rt} when discussing synthetic spectra from our radiative transfer simulations.

\begin{figure}
\centering
\includegraphics[width=\hsize]{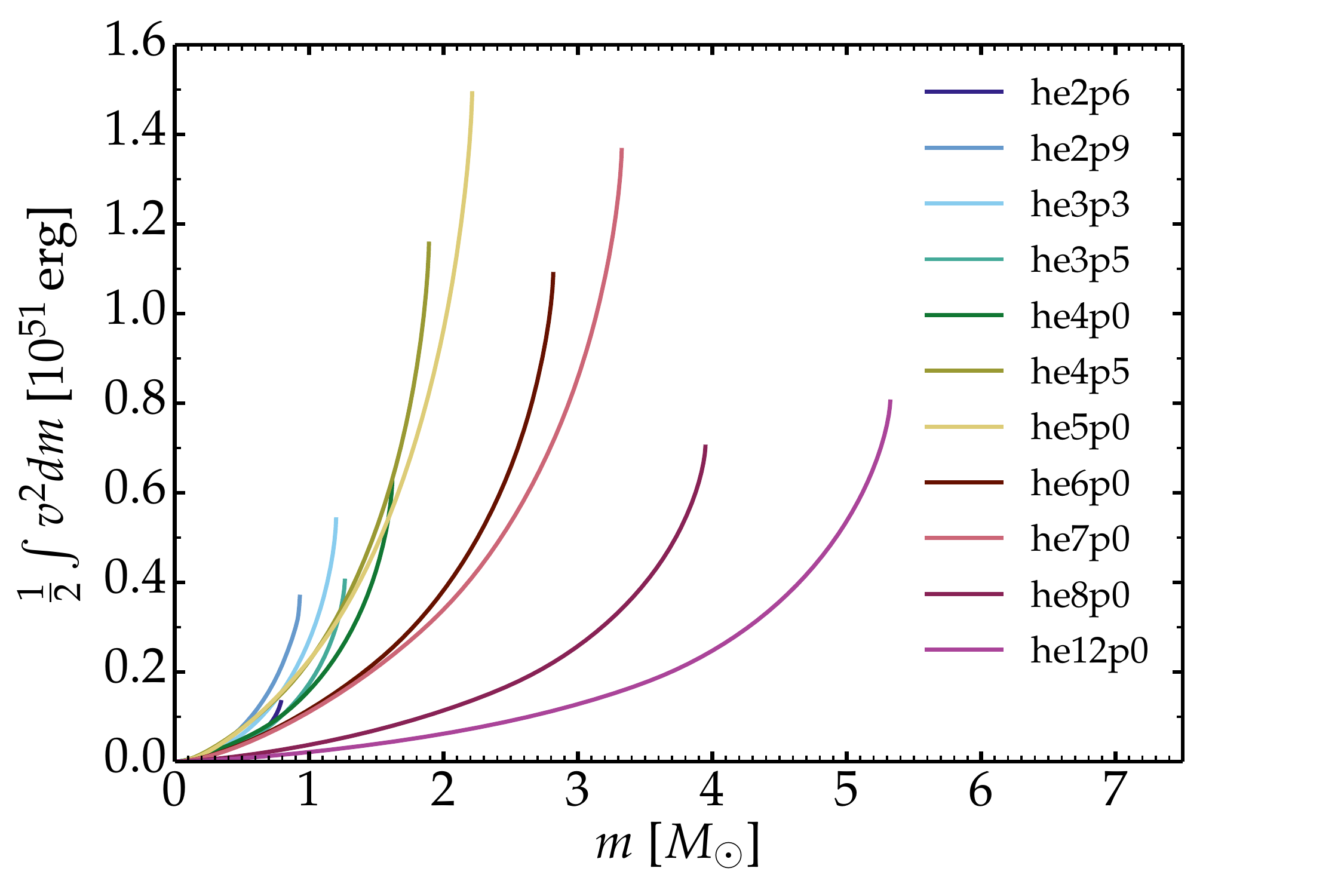}
\caption{Cumulative ejecta kinetic energy versus initial helium-star mass for the explosion models of \citet{ertl_ibc_20}. The integral is carried out from the center of the ejecta outwards. Moderate ejecta kinetic energies of order a few 10$^{50}$\,erg are found in the lower mass helium-star models, which also have ejecta masses $\lesssim$\,1\,\msun.
\label{fig_ekin}
}
\end{figure}

Another interesting aspect concerns the typical explosion energies estimated for the helium-star progenitors of Type Ibc SNe. Figure~\ref{fig_ekin} shows the cumulative ejecta kinetic energy versus mass for the helium-star explosion models of \citet{ertl_ibc_20}, which are based on the helium-star progenitor models of \citet{woosley_he_19} evolved with a nominal mass loss rate. As discussed in the preceding paragraphs and in Fig.~\ref{fig_he}, helium represents at least 50\,\% of the ejecta mass up to the helium-star model with an initial mass of 5.0\,\msun, and then becomes sub-dominant at higher mass. It is also most abundant in the outer ejecta, and progressively replaced by atoms of increasing atomic mass as one progresses inwards in the ejecta. These helium-star models yield ejecta kinetic energies between about 0.5 and $1.5 \times 10^{51}$\,erg. Thus there is a lot of momentum and kinetic energy stored in these ejecta. This may conflict with the relatively narrow lines (suggestive of an expansion velocity of about 2000\,\kms) inferred from the spectra of many Type Ibn SNe \citep{pasto_ibn_08}.

\begin{table*}
\caption{Summary of interaction configurations simulated with \heracles\ together with properties
    of the CDS and the model radiation.
\label{tab_her}
}
\begin{center}
\begin{tabular}{
l@{\hspace{2mm}}|
c@{\hspace{2mm}}c@{\hspace{2mm}}c@{\hspace{2mm}}|
c@{\hspace{2mm}}c@{\hspace{2mm}}|
c@{\hspace{2mm}}c@{\hspace{2mm}}|
c@{\hspace{2mm}}c@{\hspace{2mm}}c@{\hspace{2mm}}
}
\hline
    Model &    \multicolumn{3}{|c}{Explosion}   &     \multicolumn{2}{|c}{CSM: Wind}  &     \multicolumn{2}{|c|}{CDS} & \multicolumn{3}{|c}{Light curve} \\
\hline
    &       $E_{\rm kin}$ &  $M_{\rm ej}$  & $M$(\nifs) &  \mdot  &    \vinf    &   $V$(CDS)  &   $M$(CDS) & $t_{\rm peak}$ & $L_{\rm peak}$ & $\int L dt$ \\
\hline
    &        [10$^{49}$\,erg]    &    [\msun] &    [\msun]    &  [\msunyr]  & [1000\kms]    &    [1000\,\kms] &   [\msun] & [d] & [10$^{42}$\,\ergs]& [10$^{49}$\,erg]  \\
\hline
W1              &   7.5      &           1.49   & 0.08  &   0.001  &   1.0  &   6.37  &       0.0051   & 29.8    &     1.58 &   0.45      \\  
W2              &   7.5      &           1.49   & 0.08  &   0.01   &   1.0  &   4.99  &       0.0201   & 28.8    &     1.73 &   0.53      \\  
W3              &   7.5      &           1.49   & 0.08  &   0.1    &   1.0  &   3.57  &       0.0974   &  1.1    &     3.43 &   0.78      \\  
W4              &   75.0     &           1.49   & 0.08  &   0.001  &   1.0  &  18.98  &       0.0086   & 11.7    &     4.45 &   0.96      \\  
W5              &   75.0     &           1.49   & 0.08  &   0.01   &   1.0  &  13.65  &       0.0424   &  0.3    &    23.73 &   2.59      \\  
W6              &   75.0     &           1.49   & 0.08  &   0.1    &   1.0  &   8.83  &       0.2523   &  1.1    &    91.73 &   8.90      \\  
W7              &   7.5      &           0.15   & 0.008  &   0.001  &   1.0  &  13.71  &       0.0043   &  0.2    &     4.24 &   0.31      \\  
W8              &   7.5      &           0.15   & 0.008  &   0.01   &   1.0  &   8.84  &       0.0254   &  0.3    &    12.26 &   0.95      \\  
W9              &   7.5      &           0.15   & 0.008  &   0.1    &   1.0  &   5.59  &       0.1168   &  1.0    &    30.77 &   2.52      \\  
\hline
\hline
Model   &    \multicolumn{3}{|c}{Explosion}     &     \multicolumn{2}{|c}{CSM: Ejecta}  &     \multicolumn{2}{|c|}{CDS} & \multicolumn{3}{|c}{Light curve} \\
\hline
    &     $E_{\rm kin}$ &  $M_{\rm ej}$  & $M$(\nifs) & $E_{\rm kin}$ &  $M_{\rm ej}$   &  $V$(CDS)  &   $M$(CDS)   & $t_{\rm peak}$ & $L_{\rm peak}$ & $\int L dt$ \\
\hline
    &          [10$^{49}$\,erg]    &    [\msun]  &    [\msun]    &    [10$^{49}$\,erg]     &   [\msun]   &    [1000\,\kms] &   [\msun]  & [d]  & [10$^{42}$\,\ergs] & [10$^{49}$\,erg]\\
\hline
E1          &         7.5            &       1.49  & 0.08     &    0.01       &    1.0       &      1.67    &     1.4761  & 9.3  &   39.2 &  2.48 \\
E2          &         7.5            &       1.49  & 0.08     &    1.0        &    1.0       &      1.96    &     1.3637  & 10.8 &   13.3 &  1.45 \\
E3          &         75.0           &       1.49  & 0.08     &    0.01       &    1.0       &      4.56    &     1.8138  & 8.9  &  460.6 & 29.1  \\
E4          &         75.0           &       1.49  & 0.08     &    1.0        &    1.0       &      4.91    &     1.7519  & 9.1  &  323.8 & 23.1  \\
E5          &         7.5            &       0.15  & 0.008    &    0.01       &    1.0       &      1.37    &     1.0127  & 11.4 &   41.8 &  5.7  \\
E6          &         7.5            &       0.15  & 0.008    &    1.0        &    1.0       &      2.08    &     0.8853  & 10.7 &   33.3 &  4.2  \\
\hline
\end{tabular}
\end{center}
{\bf Notes:} The properties of the CDS are given at 40\,d after the onset of the interaction. At this time, most of the CSM has been swept-up so the
    CDS properties are no longer changing. For the time-integrated bolometric luminosity, the integral covers from the time of the first record of
    radiation at the outer boundary until the end of the simulations at 50\,d. This covers the main part of the high-brightness phase.
\end{table*}

The composition and mass of the CSM produced in hydrogen-deficient progenitors therefore departs significantly from their hydrogen-rich counterparts and Type IIn SNe. Massive stars generally have little material with a composition dominated by helium. Having a larger escape speed, the characteristic wind velocities (of order 1000\,\kms) from hydrogen-deficient stars are much larger than those in hydrogen-rich supergiants (of order 100\,\kms). Together, these properties suggest that the shells involved in the interaction may typically have lower densities than in Type IIn SNe. Helium is also harder to ionize than hydrogen (24.6\,eV for He\one\ compared to 13.6\,eV for H\one) so that under the conditions prevalent in Type Ibn SNe, helium will likely be partially ionized. This implies a reduction in electron scattering opacity by a factor of 5--10 (and typically an even greater reduction in optical depth) relative to Type IIn SNe. In the interaction scenario, all these properties suggest that Type Ibn SNe should have much faster evolving light curves.

\section{Ejecta/CSM interaction: Insights from radiation-hydrodynamics simulations}
\label{sect_rhd}

 In general, Type Ibn SNe exhibit a fast rise to a bolometric maximum of order 10$^{43}$\,\ergs\ and with a high brightness phase that lasts typically 10\,d -- the SN brightness quickly drops after maximum \citep{pasto_ibn_08,pasto_14av_16,hosseinzadeh_ibn_17}. The time-integrated bolometric luminosity is thus of the order of 10$^{49}$\,erg. Most Type Ibn SNe never show broad lines typical of standard Type Ibc SN ejecta, which is intriguing.\footnote{We note, however, that the Ca\two\ near-infrared triplet exhibits a large width at late times in some Type Ibn SNe like 2006jc \citep{pasto_06jc_07}. Unfortunately, we have not  been able to elucidate this property.} If the total kinetic energy of the inner shell before interaction was large, a strong deceleration would have resulted in a much larger  time-integrated bolometric luminosity, of at least several 10$^{50}$\,erg, which is not the case. If both shells moved fast and resulted in a weak interaction, the deceleration could be small, but then the lines would be broad throughout the evolution, which is not the case. So, the initial ejecta kinetic energy of the inner shell must be small,  atypical of a standard explosion. Further, the lack of hydrogen in Type Ibn SNe, the dominance of He\one\ lines at all times, and the limited helium mass (free of hydrogen) in massive stars at around 1\,\msun\ (see Sect.~\ref{sect_he}) suggest that the total mass involved in these interactions is small and of the order of a solar mass. Finally, there is documented evidence  that in some Type Ibn SNe the outer shell was produced a few years before  the progenitor star exploded (e.g., SN\,2006jc; \citealt{pasto_06jc_07}), which suggests the interaction  takes place at large distances.  Guided by these characteristics, the aim of this section is to clarify what interaction configurations are compatible with Type Ibn SNe.

 It is worthwhile to clarify what is meant by ejecta and wind, which we use in turn to parametrize the CSM. An ejecta is produced by a very short but large energy deposition at some depth in the stellar interior. This leads to the formation of a shock and its outward propagation through the star eventually causes the unbinding of a fraction or all of the overlying layers. The timescale for the energy injection is extremely small compared to the expansion time scale of the ejecta. In contrast, a wind is driven by the continuous supply of energy or momentum at and above the stellar surface. It is a surface phenomenon that peels off the outermost layer of the star. In the case of a radiation-driven wind, a large photon luminosity has to be provided for the entire duration of the wind phase. Hence, the phenomena at the origin of an ejecta or a wind are entirely different.

 \subsection{Numerical setup}

We performed  1D multi-group radiation hydrodynamics simulations for a variety of interaction configurations involving an inner shell and an external shell. The inner shell was assumed to result from an explosion, most likely following core collapse. For the outer shell, we considered two possible scenarios. The material was either explosively ejected (as would occur, for example, in a nuclear flash; \citealt{woosley_flash_80}; \citealt{dlw10a}; \citealt{WH15}), or driven out in the form of a super-Eddington wind (for example as a result of excitation by waves born in the core of the progenitor;  \citealt{quataert_shiode_12}; \citealt{fuller_rsg_17}; see also  \citet{quataert_superedd_16} and \citet{owocki_superedd_17} for the mechanisms behind such winds). The distinction between these two cases is nontrivial. In SN ejecta, the bulk of the mass lies at low velocity and the velocity is a linear function of radius. In the case of a wind, the mass is more uniformly spread in radius (following a $1/r^2$ dependence) and the velocity is constant with radius (for a more detailed discussion of the implication of these two configurations, see discussion in \citealt{D16_2n}).

The 1D multi-group radiation hydrodynamics simulations were carried out with \heracles\ \citep{gonzalez_heracles_07,vaytet_mg_11}, using the approach discussed in \citet{D15_2n}. An ideal-gas equation of state was used for simplicity, and also because the bulk of the internal energy in explosions and interactions is stored in radiation. For the present simulations on Type Ibn SNe, we considered only hydrogen-free material. The configurations for the inner and outer shells were taken from the helium-star progenitor and explosion models of \citet{dessart_snibc_20} -- these models are very similar to those produced by \citet{woosley_he_19} and \citet{ertl_ibc_20}. The complex composition of these models was simplified so that \heracles\ would only follow helium, oxygen, silicon, and iron (with a suitable renormalization of the sum of mass fractions to unity) in order to capture the basic chemical stratification of these ejecta. This was done for the purpose of using the adequate, composition-dependent, opacities at each location of the grid. In practice, opacities were calculated as a function of composition, density, temperature and energy group (for details, see \citealt{D15_2n}). We used eight energy groups positioned at strategic locations to capture the strong variation in absorptive opacity with wavelength. One group covers the  Lyman continuum, two groups sample the Balmer continuum, two other groups cover the Paschen continuum, and finally three groups sample the Brackett continuum and beyond. We include radioactive decay power from \nifs\ and \cofs\ and assume that this power is deposited at the site of emission.

For these radiation-hydrodynamics simulations, we were first interested in the bolometric light curve properties, in particular the rise time to peak, the peak luminosity, and the time-integrated bolometric luminosity over the high-luminosity phase. We also wanted to study the properties of the CDS that inevitably forms at the junction of the inner shell and outer shells, composed of swept-up CSM and decelerated ejecta material. The diversity of light curve and dynamical properties was generated by varying the properties of the inner and outer shells.  We studied the configuration of a standard-energy core-collapse SN explosion of a Wolf-Rayet star interacting with a dense wind, but also explored other configurations that departed significantly from this.

The default ejecta model for the inner shell is the solar-metallicity model he4 from \citet{dessart_snibc_20} at 1\,d after explosion (this model is a close analog of model he4p0 from \citet{ertl_ibc_20}, but differs slightly in ejecta mass and kinetic energy -- to distinguish these two models, we use he4 to refer to the model from \citealt{dessart_snibc_20} and to he4p0 to refer to the model from \citealt{ertl_ibc_20}). We adopted the radius, velocity, density, temperature and composition from that model, including \nifs. The ejecta is hot and optically thick at that time, with a significant storage of radiation energy so we start with the initial temperature of this ejecta at  that time. For the outer shell, we assumed a composition identical to the outermost point in the he4 model (i.e., in the He/N shell), which is essentially pure helium and metals at their solar metallicity value (there is no \nifs\ in the outer shell). For the outer shell, an analytical description of the fluid variables was used for both the ejecta case (for details, see \citealt{dessart_audit_rhd_3d_19}), or for the wind case (assuming a wind velocity $V_\infty$ and a density $\rho = \dot{M} / 4 \pi R^2 V_\infty$, where $\dot{M}$ is the wind mass loss rate and $R$ is the radius). The outer shell material was assumed to be cold initially, with a temperature of 2000\,K.

 We joined the inner and outer shells at a radius $R_{\rm t}$ of about 10$^{14}$\,cm (see Fig.~\ref{fig_prop_init}). This corresponds to an age for the inner shell of about one day after core collapse, while the outer shell was produced months to years before core collapse (depending on its velocity). Different values of $R_{\rm t}$ reflect different physical evolutions for the progenitor star but do not alter much the overall trends discussed here, as long as $R_{\rm t}$ remains below a few 10$^{15}$\,cm  (for larger values of $R_{\rm t}$, the outer shell would be optically-thin, the thermalization would be weaker, the deceleration of the inner shell less efficient etc.)

\begin{figure*}
\centering
\includegraphics[width=0.495\hsize]{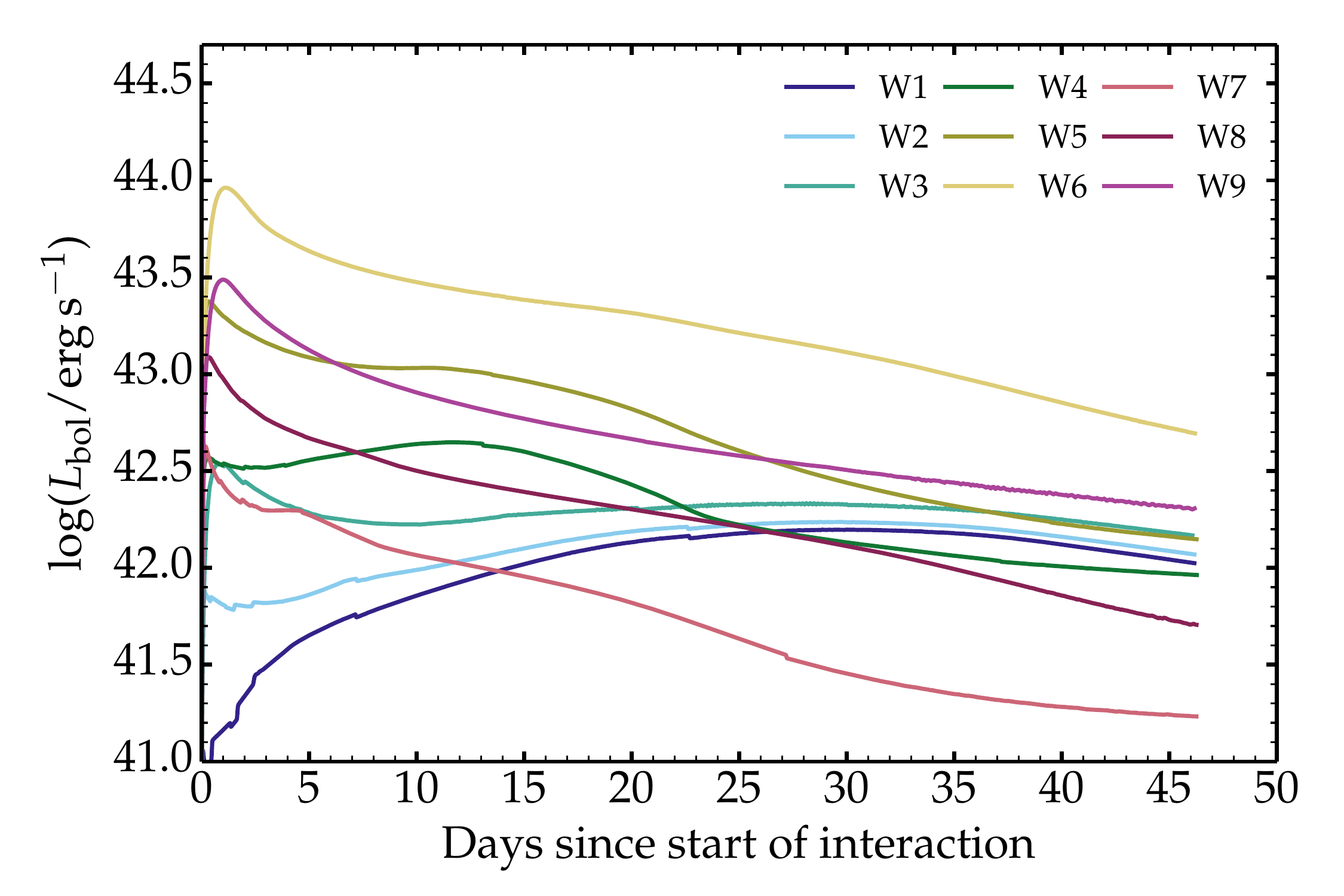}
\includegraphics[width=0.495\hsize]{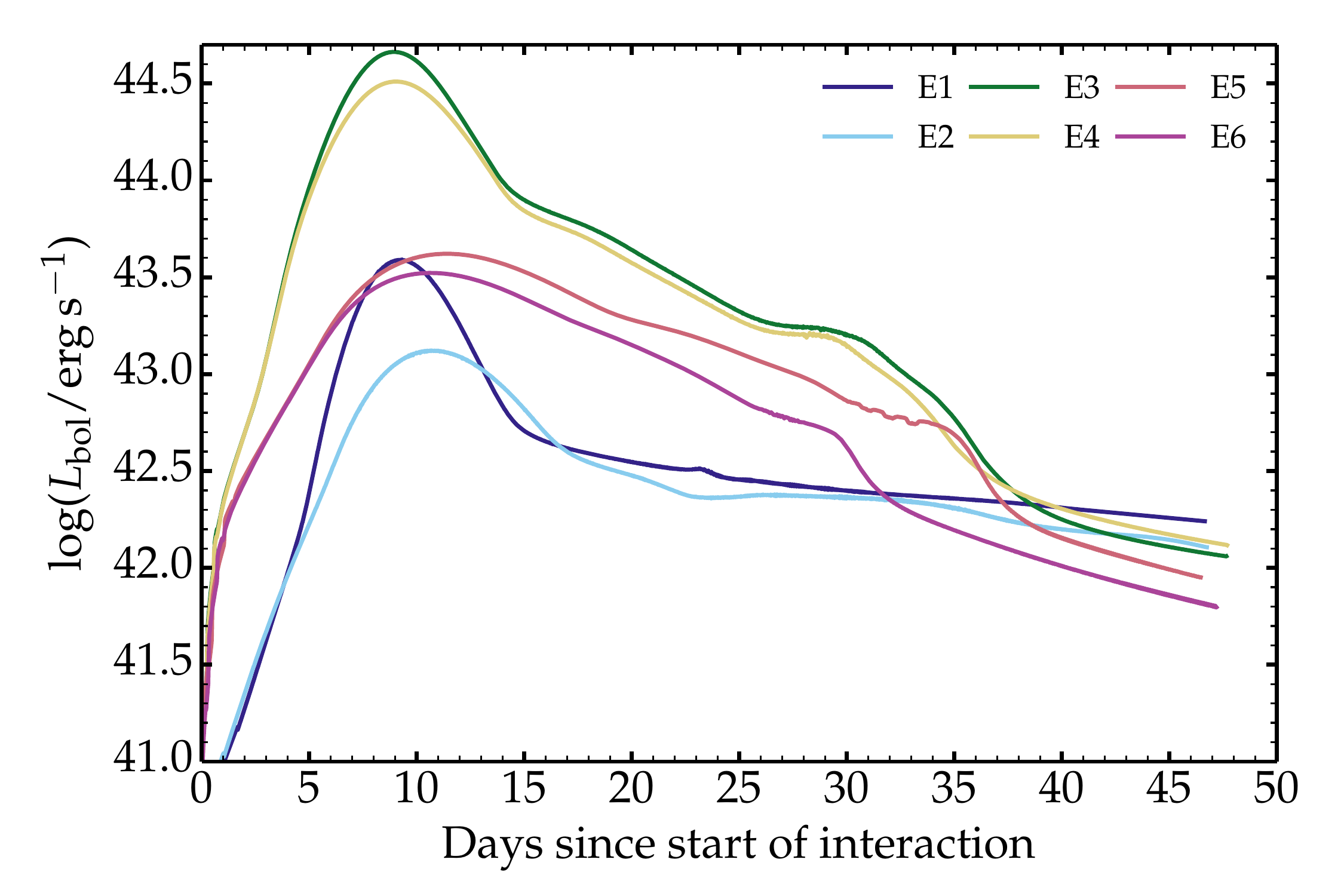}
\caption{Bolometric light curves computed with \heracles\ and for interaction configurations in which the outer shell corresponds to a dense wind (left panel; discussion in Sect.~\ref{sect_wind}) or to an ejecta (right panel; discussion in Sect.~\ref{sect_ejecta}). Interaction configurations are summarized in Table~\ref{tab_her}.
\label{fig_lbol_her}
}
\end{figure*}

For the inner shell, the default kinetic energy is $E_{\rm kin} = 7 \times 10^{50}$\,erg, $M_{\rm ej} =$\,1.49\,\msun, and  $M(^{56}$Ni$)=$\,0.08\,\msun. Variants were produced in which we scaled the velocity by a factor $\sqrt{0.1}$ (corresponding to a scaling of $E_{\rm kin}$ by 0.1) and others in which we scaled the density by a factor of 0.1 (corresponding to a scaling of both $E_{\rm kin}$, $M_{\rm ej}$, and $M(^{56}$Ni) by 0.1). These scaled variants are not fully consistent but they are useful to explore the outcome of interactions involving a less energetic or a less massive inner ejecta. For the outer shell, we considered either an ejecta of 1\,\msun\ with $E_{\rm kin}$ of 10$^{47}$ or 10$^{49}$\,erg (models E1 to E6), or winds with $V_\infty=$\,1000\,\kms\ and wind mass loss rates of 0.001, 0.01, or 0.1\,\msunyr\ (models W1 to W9).  Table~\ref{tab_her} presents a summary of the \heracles\ simulations, both for the initial interaction configurations and for the resulting light curve and CDS properties. We show the initial radial profile of the velocity, density, and temperature for all interaction configurations in Fig~\ref{fig_prop_init}.

 \subsection{Results}

 The resulting dynamical and radiative properties reflect how the kinetic energy stored in the outer parts of the inner shell is absorbed by the outer shell. In turn, the optical depth of the outer shell sets the typical diffusion time over which the bulk of this power is radiated (this timescale is also comparable to the rise time to maximum). When the configuration becomes optically thin, the model luminosity is equal to the shock luminosity $L_{\rm shock} = 2 \pi R^2 \rho V_{\rm shock}^3$ (here, $R$ is the radius of the shock, $\rho$ is the CSM density ahead of the shock, and $V_{\rm shock}$ if the shock velocity) together with the contribution from radioactive decay power.

In the next two sections, we present some results from radiation-hydrodynamics simulations. We do not intend in this work to be quantitative and to propose a specific model for any specific observation. Rather, we explore the systematics for various configurations, and give special attention to the salient features of transient light curves, namely the rise time to peak, the peak luminosity, and the time-integrated bolometric luminosity over the high-luminosity phase. Detailed comparisons of models and observations are deferred to a future study.

 \subsubsection{Outer shell: the wind case}
\label{sect_wind}

 The left panel of Fig.~\ref{fig_lbol_her} shows the bolometric light curves for the interaction of an ejecta with a wind. The wind mass loss rates cover the range from 0.001 to 0.1\,\msunyr\ and have at best a moderate optical depth. If the shock power is small relative to the decay power (models W1, W2, W3 with a scaled velocity), the rise to maximum and the luminosity at maximum are essentially unchanged from the case with no wind (for a \nifs\ mass of 0.08\,\msun, the peak bolometric luminosity is at about $10^{42.2}$\,\ergs).\footnote{The rise time is  longer than obtained in \citet{dessart_snibc_20}, in spite of the neglect of line opacity in \heracles. This longer rise time occurs because we have scaled down the velocity.} However, the shock power boosts the model brightness before maximum, when it is otherwise low in such ejecta. This configuration does not directly apply to Type Ibn SNe like 2006jc but is important for early-time properties of Type Ibc SNe. It shows that wind interaction can produce a bolometric light curve analogous to that resulting from a progenitor star with an extended radius (see, for example, \citealt{bersten_etal_12_11dh}; \citealt{piro_15}; \citealt{d18_ext_ccsn})

 If the shock power is large and typically larger than the contribution from decay power, the peak luminosity is dominated by interaction power, and the rise time is set by the diffusion time through the CSM, which is of the order of a day (a longer rise time in the wind/CSM case requires a CSM closer to the star or a greater wind mass loss rate). One exception is model W4 in which the first peak due to interaction power is of comparable strength to the second peak, which is due to both decay power and interaction power. This is reminiscent of the double-peak light curve observed in SN\,2011hw \citep{pasto_ibn_trans_15}. The modest boost due to interaction places model W4 among the brightest Type Ibc SNe that are neither broad lined nor associated with a GRB \citep{drout_11_ibc,prentice_ibc_16,dessart_snibc_20,woosley_ibc_21}. Hence, wind interaction may be the currently missing power source for Type Ibc SNe  observed in this range of peak luminosities. The impact on line profiles would be limited to differences in absorption and emission at the highest velocities, so that it may not be so evident from spectra taken around the time of the \nifs-powered maximum or later.

 For the simulation with the largest ejecta kinetic energy and the denser wind (model W6), the light curve rises in about one day to a bolometric maximum of $\sim$\,10$^{44}$\,\ergs, followed by a steady decline until late times. The contribution from decay power is subdominant and leaves no clear signature on the light curve. Throughout most of this light curve evolution, the shock power is unaffected by optical depth effects, i.e. the instantaneous shock power sets the bolometric luminosity at all times.

 Of critical importance here is to study the velocity of the dense shell that forms in the interaction. For models with a standard-energy ejecta, the asymptotic velocity of the CDS drops from $\sim$\,19000 (W4) to $\sim$\,14000 (W5) and $\sim$\,9000\,\kms\ (model W6), and the mass in the CDS increases, in the same order, from $\sim$\,0.01 to $\sim$\,0.25\,\msun. In these simulations, the deceleration of the ejecta by the CSM is small. The CDS  contains little mass and moves very fast, much faster than inferred from observed Type Ibn SNe. In the absence of \nifs, the model radiation would arise exclusively from the fast moving dense shell and one would observe very broad lines. In the presence of \nifs, a significant fraction of the model radiation would arise from the inner ejecta and one would thus expect narrower lines but from metal species typical of the inner ejecta composition (hence not helium). Both options appear incompatible with the observations of Type Ibn SNe.

 In models in which we scaled down the density or the velocity, the deceleration by the CSM is much greater. In model W3 in which we scaled the velocity, the CDS moves at about 3500\,\kms\ and contains about 0.1\,\msun. However the model luminosity is hardly affected by shock power, and thus much smaller than in garden-variety Type Ibn SNe.

  \subsubsection{Outer shell: the ejecta case}
\label{sect_ejecta}

The right panel of Fig.~\ref{fig_lbol_her} shows the bolometric light curves for the interaction of an ejecta with another ejecta --- similar light curves have been shown in a similar context by \citet{woosley_ibc_21}. Out of a large parameter space, we selected configurations that would seem suitable for Type Ibn SNe. Since the representative helium mass in helium-star progenitors is about 1\,\msun, we chose an outer shell of 1\,\msun. To ensure interaction, the outer shell is given a modest representative velocity $V_{\rm m} = \sqrt{2E_{\rm kin}/M_{\rm ej}}$ of 100 or 1000\,\kms\ ($E_{\rm kin}$ of 10$^{47}$ or 10$^{49}$\,erg). At the onset of interaction when the inner ejecta is 1\,d old, the age of the outer shell is 750 or 24\,d, in the same order.

Because of the large CSM mass (mostly confined to the inner parts of the outer shell), the diffusion time through the outer shell is now much larger than in the wind case above. The rise time to peak is about 10\,d in all cases, but with peak luminosities that extend to even greater values. In the case of a standard energy explosion for the inner ejecta, the peak luminosity reaches $3-5\times 10^{44}$\,\ergs\ (models E3 and E4), decreasing to  $10^{42}$\,\ergs\ at 20\,d. These huge peak luminosities are comparable to that inferred for AT\,2018cow (see, for example, \citealt{margutti_18cow_19}), and arise because the CSM mass rivals the ejecta mass, leading to an efficient conversion of kinetic energy into radiation energy.

When the inner-shell velocity or density is scaled down, the peak luminosity is considerably reduced. In model E2 with an inner ejecta kinetic energy of only $7.5 \times 10^{49}$\,erg, the peak luminosity is about $10^{43}$\,\ergs\ and compatible to that inferred for SN\,2006jc. Models E5 and E6 in which the inner ejecta is ten times lighter and has ten times less \nifs, the peak luminosity is of $\sim$\,10$^{43.5}$\,\ergs\ and the interaction power is the primary power source at all times. These ejecta/CSM configurations, in which the CSM is at least as heavy as the more energetic inner shell, are comparable to the one invoked for SN\,1994W \citep{D16_2n}, which is also relevant for producing other super-luminous events like the Type IIn SN\,2006gy \citep{jerkstrand_06gy_20}.

With all six configurations, the deceleration of the inner shell is very large. The asymptotic velocity of the CDS is in the range of $\sim$\,1000 to $\sim$\,5000\,\kms. In model E2, the CDS velocity is 2000\,\kms, comparable to the expansion speeds inferred for SN\,2006jc at late times \citep{pasto_06jc_07,foley_06jc_07}. Combined with the strong deceleration, nearly all the mass available ends up in the CDS.

This exploration reveals numerous problems with the general theory for Type Ibn SNe, which is that they would arise from a standard explosion of a Wolf-Rayet star into a dense environment. In that case, a modest peak luminosity of $10^{43}$\,\ergs\ implies a very large CDS velocity and broad spectral lines. On the other hand, a modest CDS velocity of 2000\,\kms\ would require a strong ejecta deceleration which would make the peak luminosity reach values in excess of $10^{44}$\,\ergs. To match both the peak luminosity and the expansion speeds inferred from Type Ibn SNe requires another configuration.

A configuration that seems to work both for the radiation and kinematic properties of Type Ibn SNe is a low-energy low-mass ejecta ramming into a massive outer shell (as in models E1, E2, E5 or E6), ensuring the persistence of narrow lines and a modest peak luminosity. One such configuration is met with a low-mass massive star evolved in a binary system, stripped of its hydrogen-rich envelope by mass transfer, undergoing a nuclear flash or a super-wind phase in the final stages of its evolution, and ejecting the remaining material in a weak core-collapse explosion. This is an analog in a Type Ibn of what has been proposed for the Type IIn SN\,1994W \citep{D16_2n}.

\begin{figure}
\centering
\includegraphics[width=\hsize]{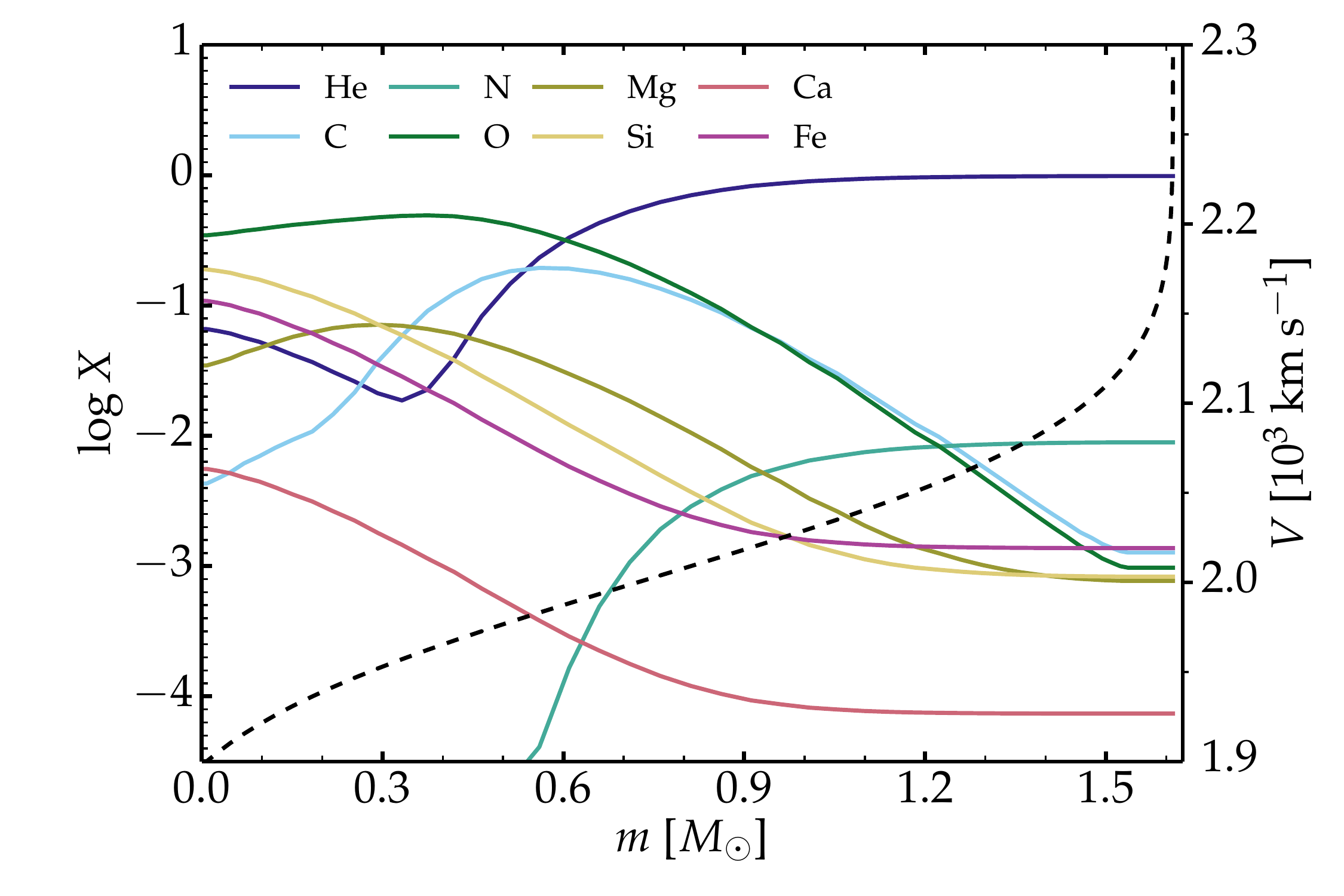}
\caption{Composition (colored solid lines) and velocity (dashed line) profiles versus Lagrangian mass for the helium-star explosion model he4p0 of \citet{ertl_ibc_20}. A boxcar mixing set to 0.06\,\mej\ has been applied four times. Our ansatz assumes that the material from the helium-star explosion and its pre-explosion mass loss has all been swept-up into a dense shell. The adopted density profile of the dense shell is given by a gaussian with a center at 2000\,\kms\ and a standard deviation of 70\,\kms.  The density in such a shell is 100 to 1000 times greater than in the inner ejecta of the he4p0 model without interaction.
\label{fig_he4p0}
}
\end{figure}

\section{Non-LTE radiative-transfer calculations for Type Ibn supernovae at nebular times}
\label{sect_rt}

\subsection{Approach and numerical setup}
\label{sect_rt_setup}

The results from the previous section indicate that a suitable configuration for Type Ibn SNe like 2006jc, 2011hw, or 2018bcc may be a 10$^{50}$\,erg ejecta with $\lesssim$\,1\,\msun\  ramming into a slowly moving 1\,\msun\ outer shell. At late times, all the mass in the system has piled up into a dense shell moving at a velocity of  order 1000\,\kms. There is little mass either below or above the dense shell, hence most of the absorption and emission arises from the dense shell. In this situation, when focusing on the low-energy radiation, it is no longer necessary to consider the global and complex structure of the interaction, and one can instead focus exclusively on the CDS.

To simplify further the approach, we may no longer distinguish between inner-shell and outer-shell material within the CDS. Weeks after the bolometric maximum, fluid instabilities will probably have had sufficient time to produce a mixed composition of both shells.\footnote{A foretaste of such instabilities in interacting SNe can be gleaned from the study of \citet{blondin_sn2n_96}. More work, in 3D, including radiation transport,  with realistic progenitors, and in the parameter space relevant for Type Ibn SNe, is needed to determine the level of mixing and clumping in the CDS.}. Furthermore, the pre-SN ejecta occurs just months to years before core collapse, so one can assume that the inner shell and the outer shell material involved in the interaction arise from the envelope above the iron core in the progenitor. For our radiative-transfer calculations, we can therefore use a helium-star explosion model, and assume that at late times, this ejecta is a good representation of the combined inner and outer shells ejected in two consecutive events. In this work, we thus use the helium-star explosion models originally presented by \citet{ertl_ibc_20}. We enforce a chemical mixing by running four times a boxcar width of $0.06 M_{\rm ej}$ through the ejecta. In our calculations, we consider solar-metallicity progenitor models with an initial helium-star mass between 2.9 and 12\,\msun\ (for details on these models, see \citealt{dessart_snibc_21}). Following our assumption for the late-time conditions, we put the total mass of  each model into the dense shell. The density structure adopted follows a Gaussian with a center at 2000\,\kms\ and a standard deviation of 70\,\kms. With this adopted structure, we rescale the density to match the ejecta mass of each helium-star explosion model -- the density in the dense shell is 100 to 1000 times greater than in the inner ejecta of the same helium-star explosion model but expanding freely without interaction. The original composition is then re-interpolated in mass space onto that new mass distribution.

We vary the age of the configuration by adopting different  radii for the CDS, here chosen to be at 2, 3, or $6 \times 10^{15}$\,cm -- an example for the model he4p0 is shown in Fig.~\ref{fig_he4p0}. For reference, a shell moving at 500\,\kms\ for two years would reach a radius of $3.15 \times 10^{15}$\,cm, which gives an idea of how far the 2004 pre-SN outburst of SN\,2006jc may have travelled \citep{pasto_06jc_07}. Because the spectrum formation occurs locally within the CDS, homologous expansion may be assumed. We no longer capture the non-monotonic velocity structure like in \citet{D15_2n}, but this allows us to run \cmfgen\ in standard, steady-state, line-blanketed mode, with a proper account of the effect of lines on the temperature and ionization structure within the CDS.

Finally, we inject power to mimic the contributions from the interaction and radioactive decay. We choose a Gaussian power profile whose volume integral yields a prescribed total power -- we explore with values of 2 to $10 \times 10^{42}$\,\ergs, which are typical of the high-luminosity phase of Type Ibn SNe like 2006jc. For the decay power, we adopt the same initial \nifs\ mass as given in each helium-star explosion model of \citet{dessart_snibc_21}, and we assume that at the time of the computation, the \nifs\ has entirely decayed into a mixture composed of 50\% \cofs\ and 50\% \fefs, as expected for a post-explosion epoch of about  three months, roughly representative of the nebular epochs that we aim to study here.\footnote{The time when the number of \cofs\ and \fefs\ nuclei are equal is given by $\tau_{\rm Co}  \log(2\tau_{\rm Co}/(\tau_{\rm Co}-\tau_{\rm Ni}))$, where $\tau_{\rm Co}$ and $\tau_{\rm Ni}$ are the lifetimes of \cofs\ and \nifs\ nuclei. This time corresponds to 86.4\,d.} When comparing to specific observations, using the inferred post-explosion time (and the adequate mixture of \cofs\ and \fefs\ -- only traces of \nifs\ would remain) would have been superior but it is not essential for this first  exploration on the basic properties of Type Ibn SNe. It would also imply rerunning a model for each observation, while there still remains much ambiguity on the actual \nifs\ mass produced in Type Ibn SNe in the first place.

\begin{figure}
\begin{center}
\includegraphics[width=\hsize]{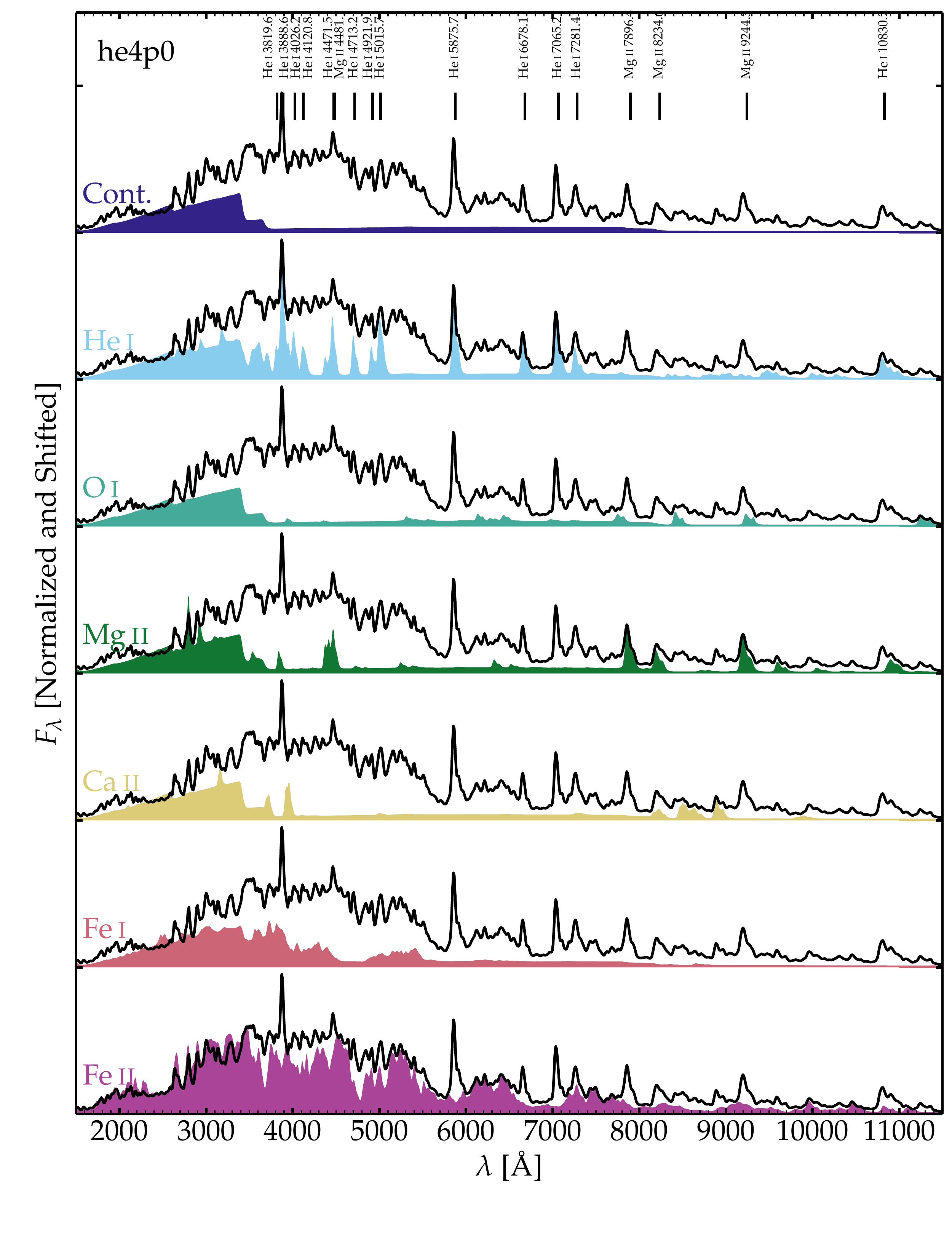}
\end{center}
\vspace{-0.8cm}
\caption{Illustration of the flux contribution from the continuum only (top row), and then He\one, O\one, Mg\two, Ca\two, Fe\one, and Fe\two\ bound-bound transitions (and continuum) to the total flux (black) in our interaction model based on the helium-star explosion model he4p0. We assume that all the model mass has been compressed into a dense shell.  The total power from the interaction is set at $2 \times 10^{42}$\,\ergs, the radius of the dense shell is at $3 \times 10^{15}$\,cm, and its velocity is 2000\,\kms. All spectra have been convolved with a gaussian kernel (with a FWHM of 23.5\,\AA -- standard deviation of 10\,\AA).
\label{fig_spec_prop}
}
\end{figure}

\begin{figure}
\begin{center}
\includegraphics[width=\hsize]{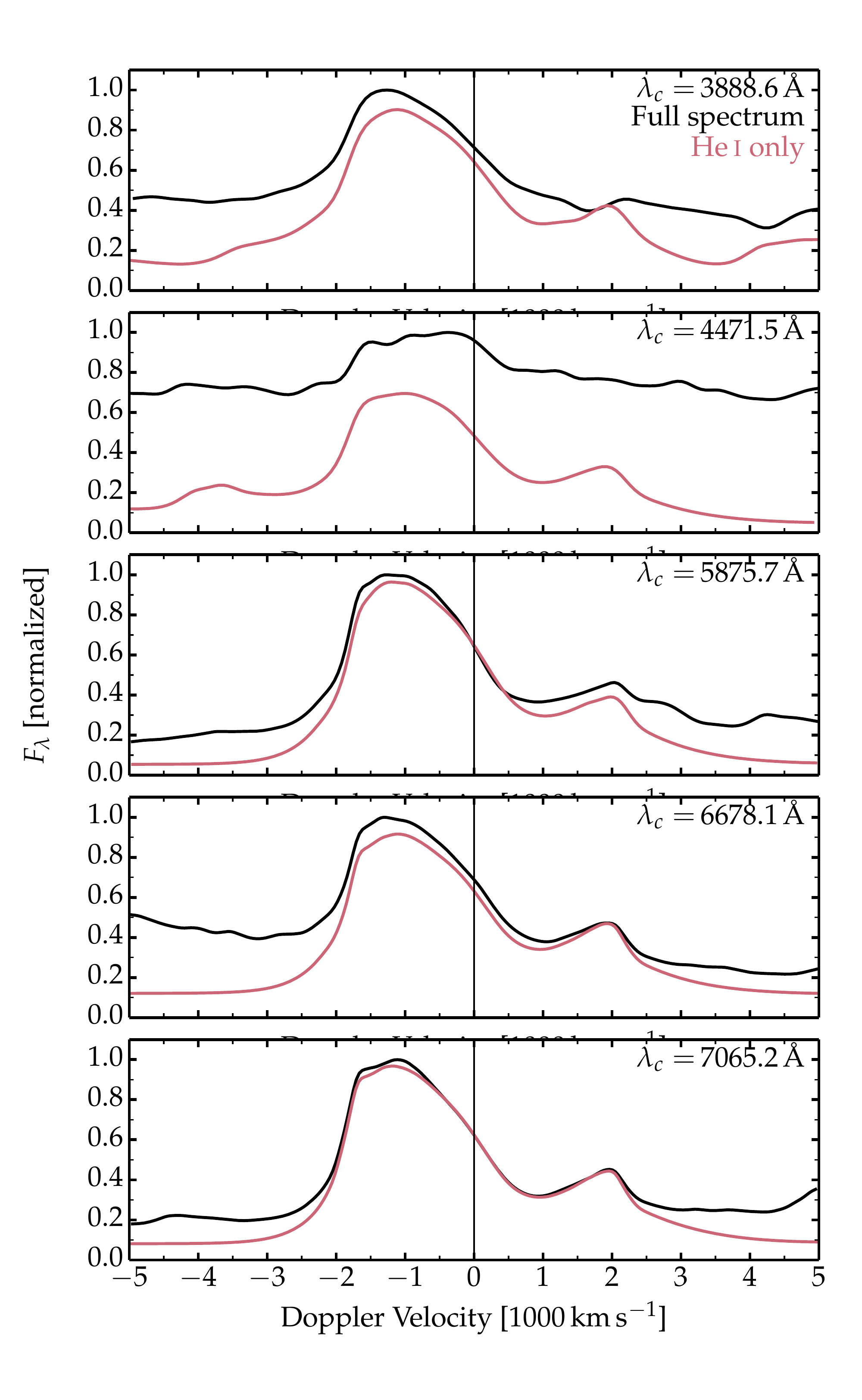}
\end{center}
\vspace{-0.2cm}
\caption{Illustration of the total flux (black) and that due to He\,\one\ bound-bound transitions (red) in the spectral regions centered on 3888.6, 4471.5, 5875.7, 6678.1, and 7065.2\,\AA. In each panel, a normalization is applied so that the total flux peaks at one -- the same normalization is then used for the He\,\one\ spectrum. A vertical offset between the two curves is typically caused by Fe\two\ emission. To erase a small level of high-frequency noise in the models, all spectra have been convolved with a gaussian kernel (with a FWHM of 4.7\,\AA -- standard deviation of 2.0\,\AA).
\label{fig_he1_asym}
}
\end{figure}

\subsection{Basic spectral properties}
\label{sect_he4p0}

Figure~\ref{fig_spec_prop} shows the resulting \cmfgen\ spectrum for a CDS composed of the entire he4p0 ejecta material (see composition and density profiles in Fig.~\ref{fig_he4p0}), centered around a radius of $3 \times 10^{15}$\,cm, with a velocity of 2000\,\kms, and subject to a total power of $2 \times 10^{42}$\,\ergs. This model exhibits a very peculiar spectrum by SN standards, but quite typical of Type Ibn SNe several weeks after bolometric maximum. With its strong Fe\two\ emission between 3000 and 5500\,\AA\, it is in fact reminiscent of the predicted nebular-phase models obtained for helium-star explosions in the mass range 2.6$-$4.0\,\msun\ (see \citealt{dessart_snibc_21}).  The bottom panel of Fig.~\ref{fig_spec_prop} shows that the model spectrum is nearly entirely an Fe\two\ spectrum (with some contribution from Fe\one), although the Fe abundance is only 1.6\% of the total mass. The spectrum also exhibits a myriad of He\one\ lines at 3889, 3965, 4026,  4471, 5016, 5876, 6678, 7065, and 7281 (note that the feature around 3900\,\AA\ is He\one\,3889\,\AA\ and not Ca\two\,H\&K), something that is not expected in nebular phase spectra of Type Ib SNe. In the red part of the optical range, the model exhibits lines of O\,\one\ (multiplets around 8446\,\AA), O\,\two\ (multiplets around 7320\,\AA), Si\two\ (around 5041--5056, 5958--5979 and 6347-6371\,\AA) and Mg\two\, (at 4481, 7896, 8234, 9218, and 9244\,\AA).

\begin{figure}
\begin{center}
\includegraphics[width=\hsize]{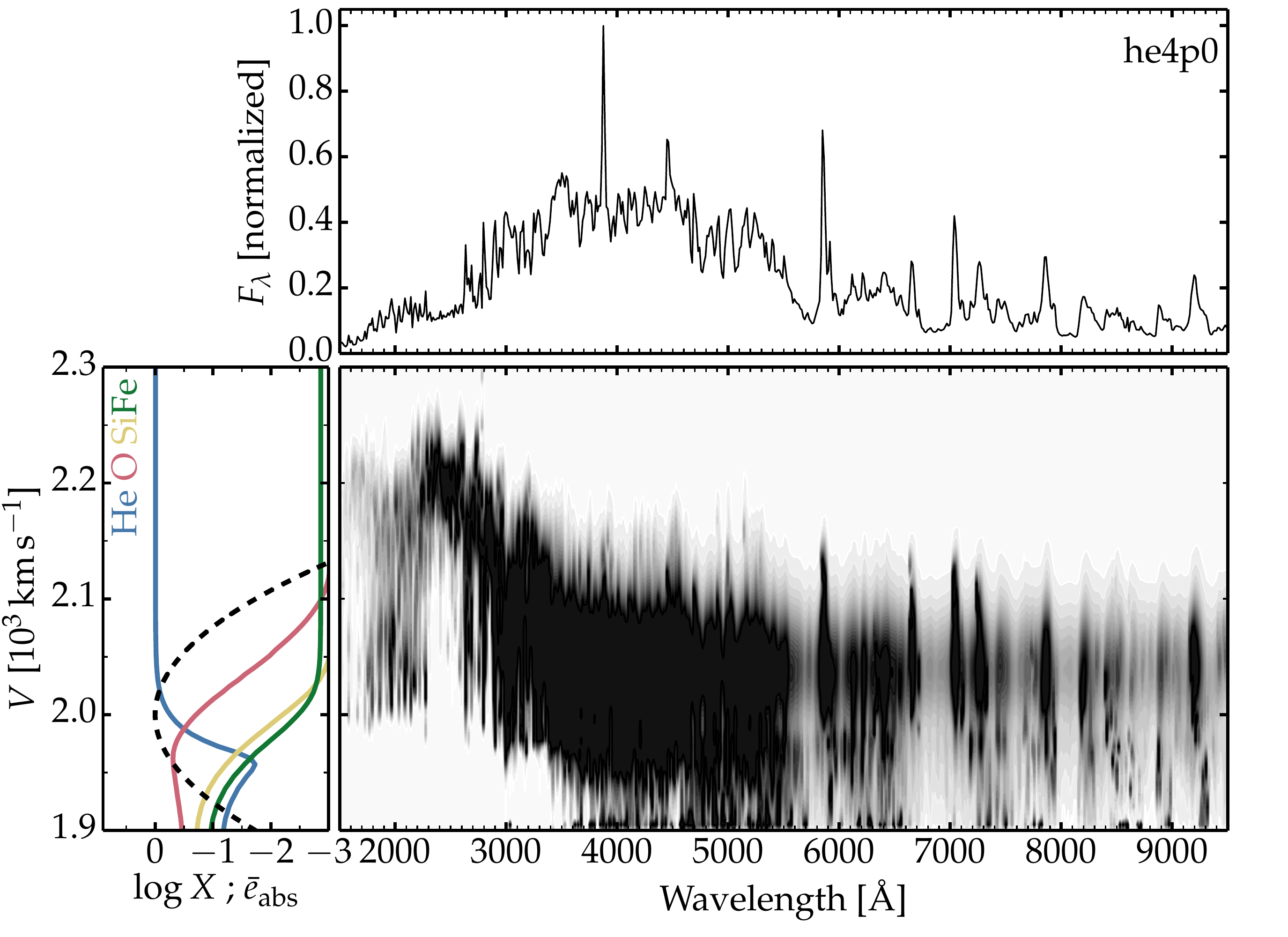}
\end{center}
\vspace{-0.2cm}
\caption{Illustration of the spectrum formation region for the model shown in Fig.~\ref{fig_spec_prop}. The top panel shows the normalized flux between 1500 and 9500\,\AA. The left panel shows the mass fraction (colored solid line) and the normalized power deposition profile (dashed) versus velocity. The central panel is a grayscale image of the observer's frame flux contribution $\partial F_{\lambda,V} /\partial V$ (the map maximum is saturated at 20\% of the true maximum to bias against the strong emission lines and better reveal the origin of the weaker emission). This map shows that a significant fraction of the Fe\two\ emission arises from regions where iron has a solar metallicity abundance.
\label{fig_dfr}
}
\end{figure}

Although a gaussian smoothing has been applied to match the typical resolution of Type Ibn SN spectra, the He\one\ line profiles predicted by the model exhibit a red deficit (Fig.~\ref{fig_he1_asym}) -- the blue emission is typically three times stronger than the red emission. This asymmetry is caused by optical depth effects. This feature may not persist in a 3D model of the interaction since the CDS would be strongly clumped, with both radial and lateral compression, thereby reducing the effective radial optical depth of the region. The blue-red asymmetry is greater for He\one\ lines in the blue part of the spectrum (e.g., He\one\,3889\,\AA) and weaker for those  in the red part of the spectrum (e.g., He\one\,5875\,\AA). This difference arises from the fact that lines are not only affected by the continuum optical depth, which is essentially gray, but also by other lines, especially in the blue part of the optical where the forest of Fe\two\  lines have a strong blanketing effect. He\one\ lines that form further out in the ejecta are less affected (see Fig.~\ref{fig_dfr}). This result is reminiscent of the differential optical depth  effects affecting H$\beta$ (which overlaps with the Fe\two\ line forest) and H$\alpha$ (which sits in an opacity hole) in models of SNe Ia that contain in their innermost ejecta layers stripped material from a companion \citep{dessart_ia_20}.

The model predicts a number of lines from O\one, Mg\two, and Ca\two.  All these metals have super-solar abundance in the mixed composition of model he4p0. O and Mg are overabundant in the ONeMg shell of the progenitor, while Ca is overabundant in the Si/S and Fe/He shells, both explosively produced. Hence, the abundance of these metals and the strengths of these O\one, Mg\two, and Ca\two\ lines are strongly model dependent (see Sect.~\ref{sect_mass}). All these lines are permitted transitions and are expected given the high density of the emitting region. This is a critical difference with standard Type Ibc SNe unaffected by interaction.

The present model has a total Rosseland-mean optical depth of 1.42 and a total electron-scattering optical depth of 0.85. A higher optical depth would have been achieved with a higher ejecta ionization: here helium (the most abundant element in this model with 0.92\,\msun) is partially ionized, O (the second most abundant element in this model with 0.31\,\msun) is once ionized, Mg is a mix of Mg$^+$ and Mg$^{2+}$, Ca is mostly Ca$^{2+}$, and Fe is present in part as Fe$^+$ but mostly Fe$^{2+}$. These ionizations explain the weak strength of O\one\ and Ca\two\ lines. The high ejecta density also inhibits the formation of \oidoub\ and \caiidoub\ that are normally seen in nebular phase spectra of Type Ibc SNe. However, the epochs and the power differ. Standard nebular-phase Type Ibc SNe are usually studied at late times of at least 150\,d past explosion, when the absorbed power, from radioactive decay, is much smaller.

\subsection{Dependence on progenitor mass and the strength of He\one\ lines}
\label{sect_mass}

In the preceding section, the CDS was filled with the ejecta from the helium-star explosion model he4p0, assuming that at the late times considered here, the ejecta and CSM had ``reunited'' following the interaction. We now use the same approach and investigate the spectral properties obtained for interactions involving ejecta and CSM from a wide range of helium-star progenitor masses.

\begin{figure}
\begin{center}
\includegraphics[width=\hsize]{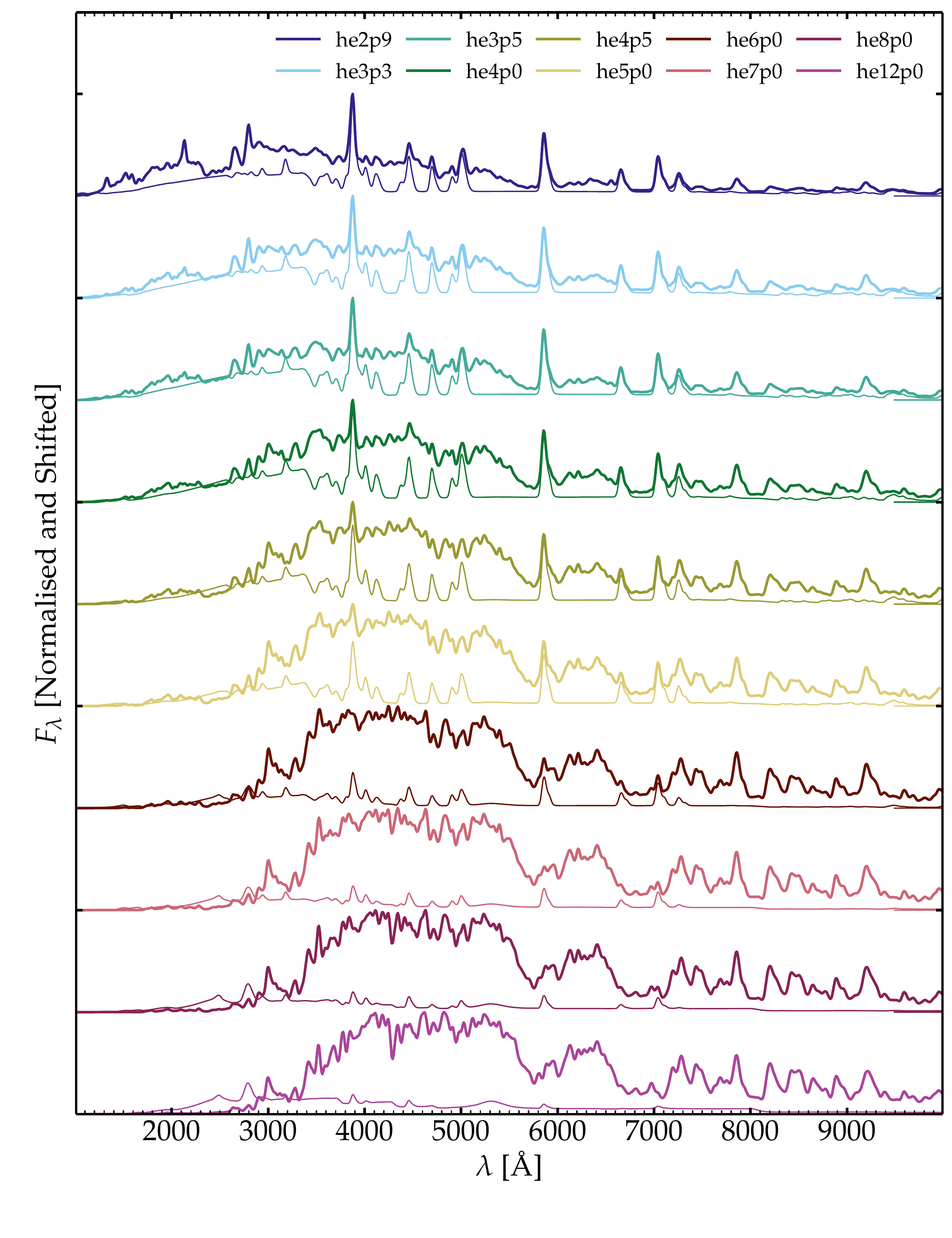}
\end{center}
\vspace{-0.8cm}
\caption{Spectral montage for models in which the CDS is composed of the material from the helium-star explosion models of \citet{ertl_ibc_20}, increasing in initial mass (he2p9 to he12p0) from top to bottom (thick line). The sequence corresponds to a decreasing fractional helium mass ($M$(He)/\mej\ varies from 83\% down to 4\%, in the same order). The CDS is at a radius of $3 \times 10^{15}$\,cm, has a velocity of 2000\,\kms, and is powered at a rate of $2 \times 10^{42}$\,\ergs. For each model, we include the spectrum corresponding to continuum processes and He\one\ bound-bound processes (thin line). He\one\ lines weaken as the helium-star initial mass is increased and the mass of helium reduced due to wind mass loss and burning to heavier elements.
\label{fig_mod_2e42_3e15}
}
\end{figure}

Figure~\ref{fig_mod_2e42_3e15} shows the UV and optical spectra for a CDS filled with the ejecta from the helium-star explosion models he2p9 to he12p0. The CDS radius is $3 \times 10^{15}$\,cm, its mean velocity is 2000\,\kms, and the deposited power is $2 \times 10^{42}$\,\ergs. As the initial helium-star mass is increased from 2.9 to 12.0\,\msun, the fractional helium mass  $M$(He)/\mej\ drops from 83\% to 4\% and the total mass (i.e., the CDS mass) increases from 0.93 to 5.32\,\msun.

Although only the yields and the total mass change between models shown in Fig.~\ref{fig_mod_2e42_3e15}, the spectra show a strong evolution. For the less massive and more helium-rich models, the spectrum is much bluer, with more flux emerging in the UV range, and most of the emission lines are associated with He\one. Where the bulk of the shock power is deposited, the CDS is hot ($\sim$\,10000\,K) and helium is nearly once ionized. This high ionization arises in part from the dominance of helium and the associated underabundance of metals, which would otherwise drive the gas to lower temperatures and lower ionization. As we progress towards a higher initial helium-star mass, the spectral energy distribution shifts to the red, He\one\ lines weaken, and lines from metals (e.g., O\one, Mg\two, or Ca\two) appear. This results from the combined effect of a lower fractional helium abundance and a lower ionization. In this sequence, we see that progenitors with an initial helium-star mass greater than 6\,\msun\ have weak He\one\ lines that are essentially impossible to discern. This suggests that Type Ibn SNe require a high fractional helium abundance of about 50\%, but not too high otherwise the spectrum is too blue (see Sect.~\ref{sect_obs}), therefore favoring a low-mass massive star progenitor in a binary and having lost its hydrogen-rich envelope through Roche-lobe overflow.

\begin{figure}
\centering
\includegraphics[width=\hsize]{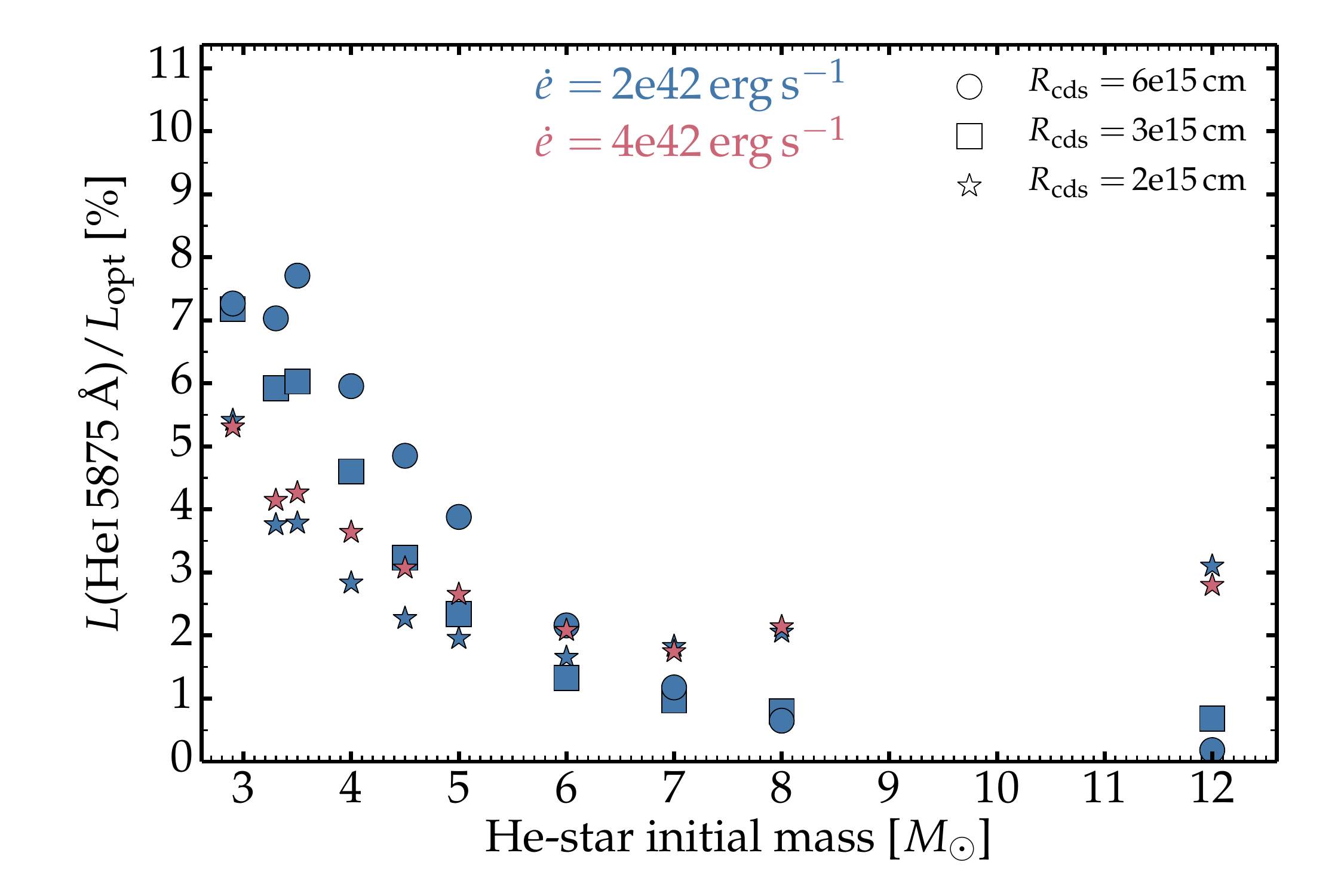}
\caption{Percentage of the optical flux (between 4000 and 8000\,\AA) that is emitted  in He\one\,5875\,\AA\ for the CDS configurations that use the helium-star models between 2.9 and 12\,\msun\ initially and for various radii and shock powers.
\label{fig_he1_5875}
}
\end{figure}

The evolution of the strength of the He\one\,5875\,\AA\ line (given as a percentage of the optical flux) is illustrated for a larger grid of models in Fig.~\ref{fig_he1_5875}. Here, we considered a CDS radius of 2, 3, or $6 \times 10^{15}$\,cm, and deposited powers of 2 and $4 \times 10^{42}$\,\ergs\ (assuming a CDS radius of $2 \times 10^{15}$\,cm).  We see that only the composition and mass of the lighter models are compatible with a strong He\one\,5875\,\AA. This supports further the notion that Type Ibn SNe are connected with weak eruptions and explosions taking place in low-mass helium stars (but massive enough to undergo core collapse).

For a smaller CDS radius, the CDS density goes up at fixed mass and the ionization goes down. A greater power yields a higher ionization and a bluer optical spectrum. The rise in temperature can also boost the continuum flux, strong in the blue, and make the CDS completely optically thick in the continuum. Conversely, for larger radii, the optical depth is lower and the spectrum is more nebular (i.e., lines sit on a weaker continuum). Clumping can reduce the ionization and influence the spectral signatures at any epoch but it does not change the optical depth. Exploring this parameter space in detail is left to a future study, but one can  envision that much diversity can arise from these different interaction configurations.

\subsection{Influence of the metal content: Iron from $^{56}$Ni decay and metallicity}
\label{sect_iron}

In this section, we present a criterion that may help distinguish Type Ibn SNe occurring in lower-mass massive stars in binaries from those that may result from pair-instability pulsations in a super massive star. The latter offer very attractive properties for Type Ibn SNe \citep{yoshida_ppisn_16,woosley_ppsn_17}, although the helium content might be too low to produce strong He\one\ lines -- they may produce a Type Icn rather than a Type Ibn SN.

\begin{figure}
\centering
\includegraphics[width=\hsize]{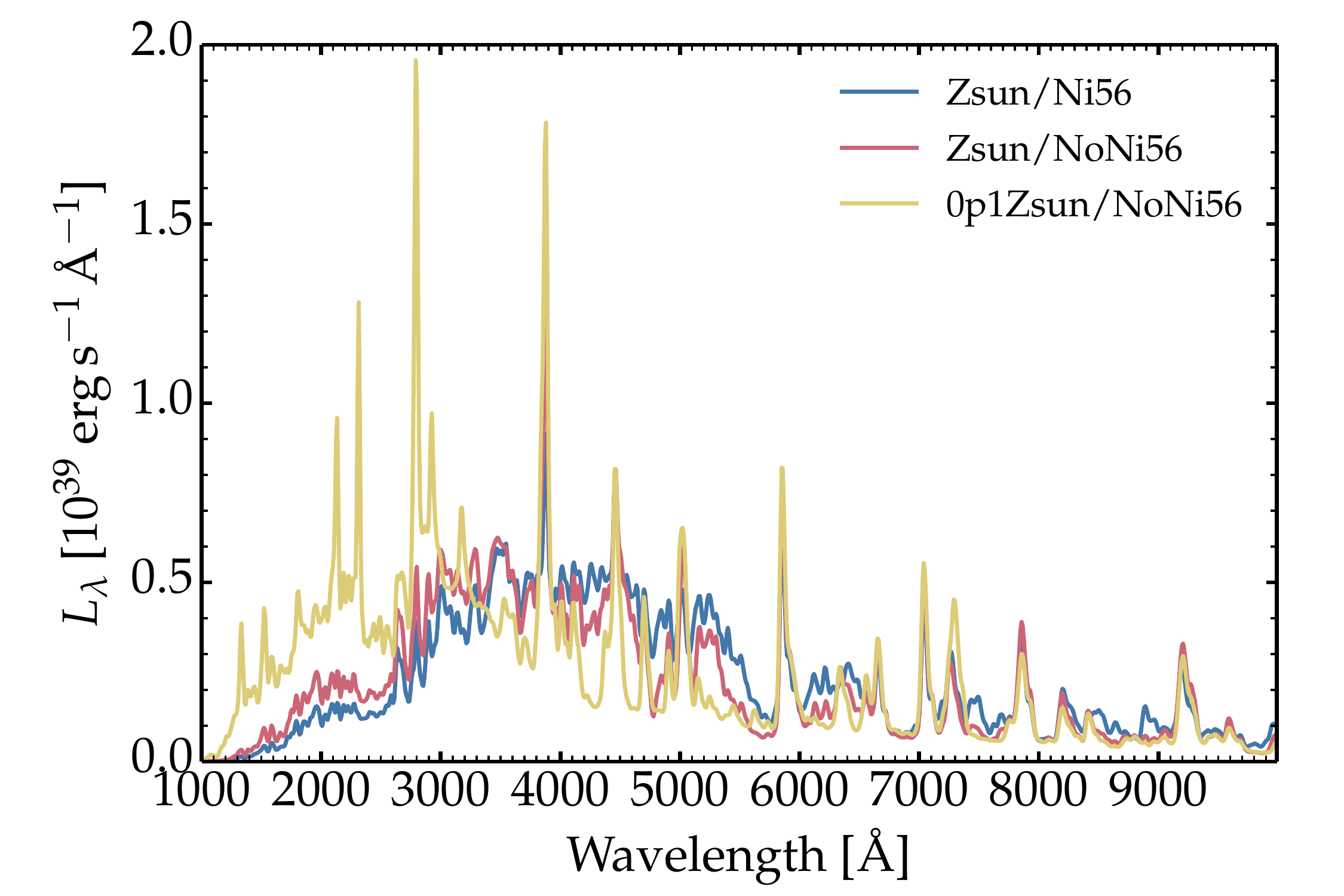}
\caption{Illustration of the influence of metals on the emergent spectrum for a Type Ibn model derived from model he4p0. The default model (see also Fig.~\ref{fig_spec_prop}) is shown in blue. The same model in which we excluded the Fe/He shell (hence no \nifs\ and associated decay products) is shown in red. Finally, the model in which we excluded the Fe/He shell and scaled by a factor of 0.1 the abundance of all metals heavier than Ar is shown in yellow.
\label{fig_ni56_Z}
}
\end{figure}

\begin{figure}
\centering
\includegraphics[width=\hsize]{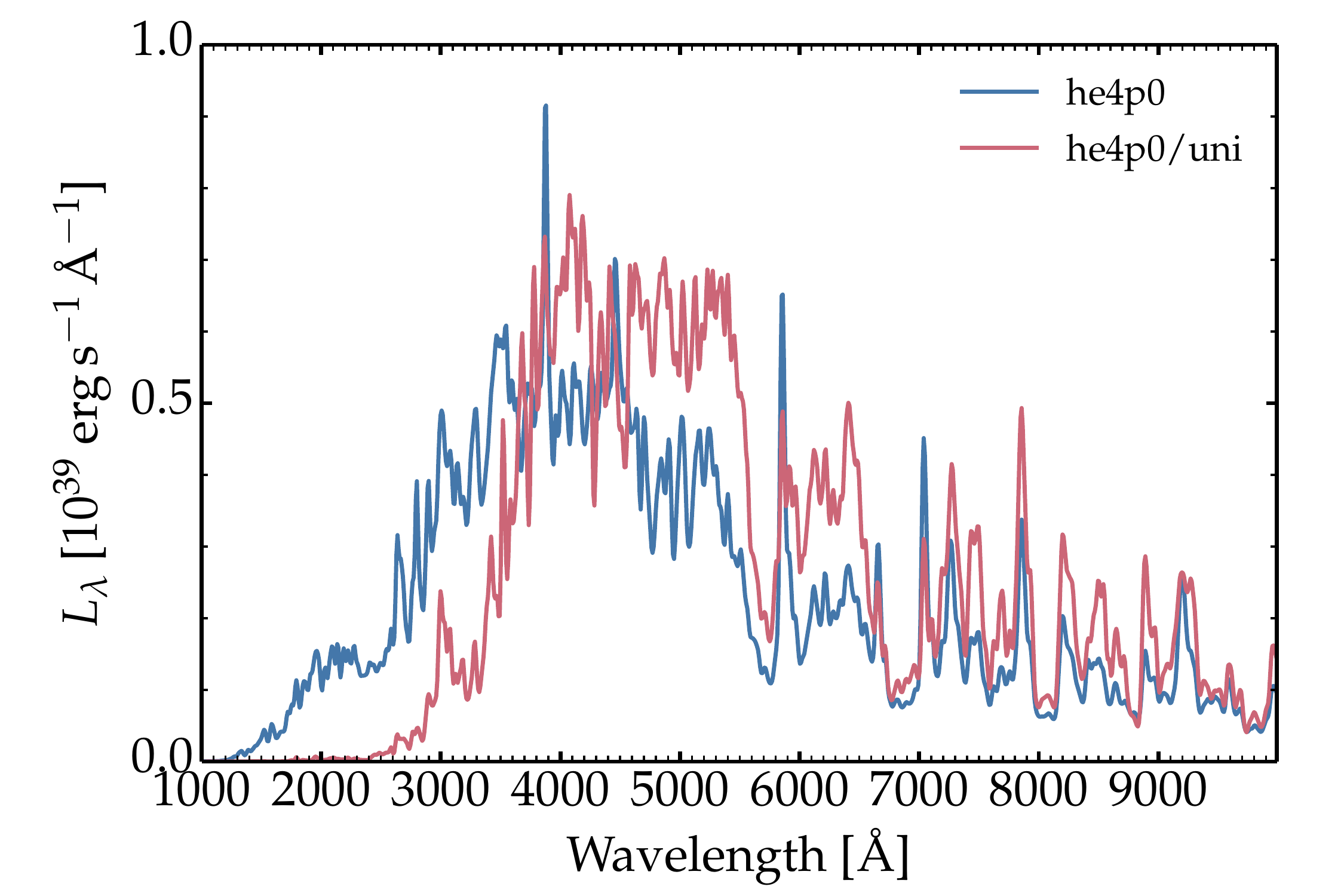}
    \caption{Spectral comparison for a CDS based on model he4p0 with mild mixing (blue; label he4p0) and the model counterpart in which the composition is homogeneous throughout (red; label he4p0/uni). This raises the iron abundance to a value that is ten times solar at every depth and the strength of Fe\two\ emission is consequently strongly enhanced in the optical but reduced in the UV.
\label{fig_unicomp}
}
\end{figure}

Two important differences may separate pulsational-pair instability in very massive stars from instabilities (such as wave excitation or a nuclear flash) in lower mass massive stars. First, the pulsational-pair instability leads to thermonuclear burning at relatively low densities where no \nifs\ is produced. Hence, at late times, these pulsations eject material that contains no \cofs\ nor any \fefs\ from the decay of \nifs\ since the \nifs\ mass is initially zero. Secondly, although strongly sensitive to the uncertain physics of wind mass loss, a low metallicity seems required for massive stars to encounter the regime of pair production and pulsation-induced ejections. The iron content of these ejecta may therefore be very low, perhaps a tenth of that expected at solar metallicity.

 Figure~\ref{fig_ni56_Z} illustrates the spectral predictions  for the default model he4p0 together with model counterparts in which the \nifs-rich shell produced in the original explosion model was excluded (this model thus contains no \fefs\ nor \cofs\ from \nifs\ decay) and another model in which, in addition, we scaled down the abundances of all metals heavier than Ar. We see that the largest impact is obtained in the model with a subsolar primordial metallicity, with an obvious shift of the spectral energy distribution to the blue, a reduced Fe\two\ emission in the optical range from 4000 to 5500\,\AA, and a strengthening of He\one\ lines. In other words, the observation of a Type Ibn SN with strong Fe\two\ emission is evidence for a solar-metallicity event, potentially in tension with a pulsational-pair instability SN.

Determining precisely the iron abundance in the ejecta is however difficult because of the unknown level of mixing within the CDS. If some \nifs\ was produced during the explosion, it could also be strongly mixed within the CDS and raise the iron abundance throughout, the more so for a large initial \nifs\ mass and a small ejecta mass. We study this possibility using model he4p0 by enhancing the mixing so that the composition becomes uniform. In model he4p0, this raises the iron abundance to 0.016 throughout the CDS, hence about ten times the solar value. The impact on the spectrum is significant, with stronger line blanketing in the blue and a boost to the Fe\two\ emission in the range from 4000 to 5500\,\AA, as well as in the range 6000 to 6800\,\AA\ (Fig.~\ref{fig_unicomp}). This hypothesis is an upper limit on the possible mixing in Type Ibn SNe and may never be realized in nature.

\begin{figure*}
\centering
\includegraphics[width=0.495\hsize]{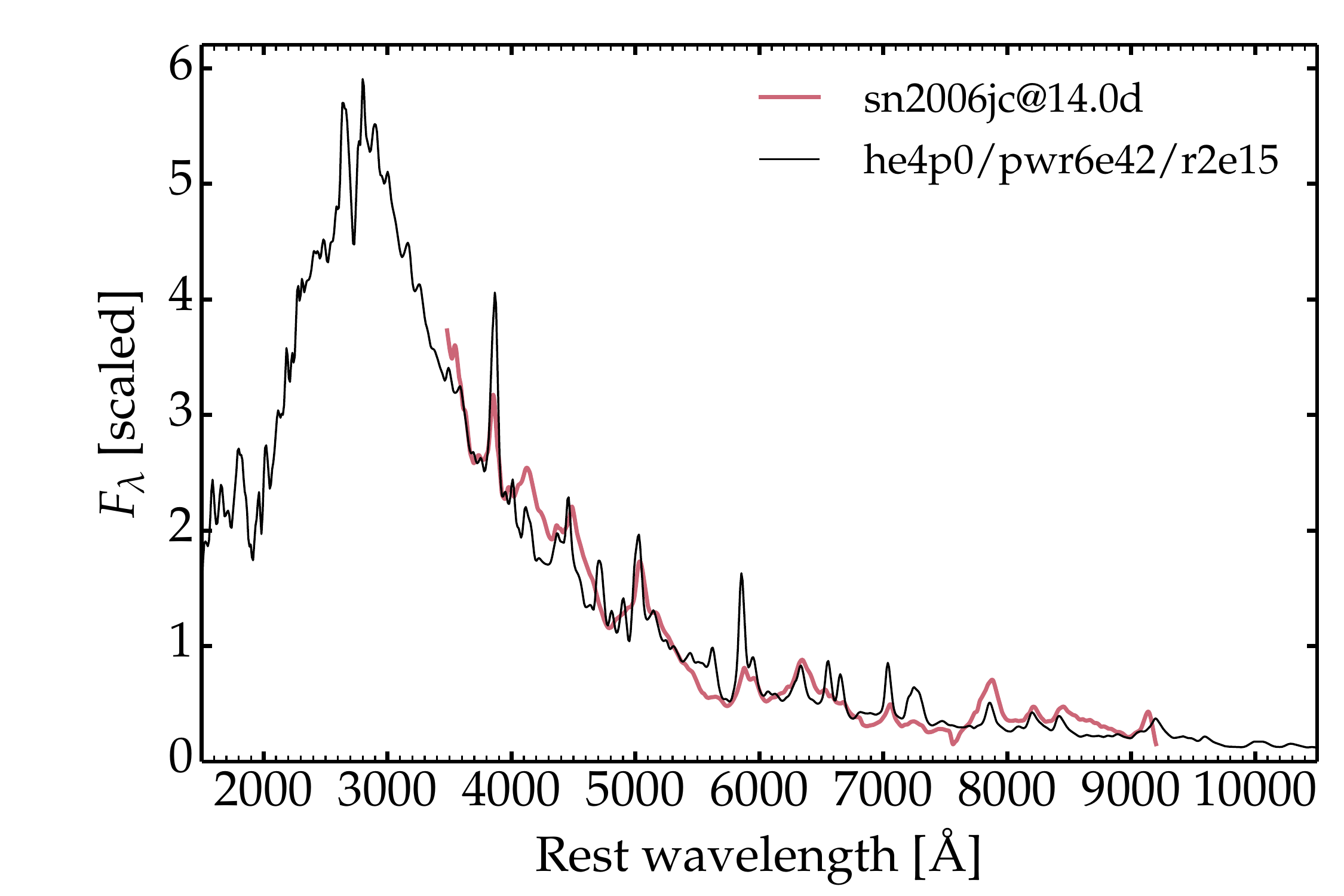}
\includegraphics[width=0.495\hsize]{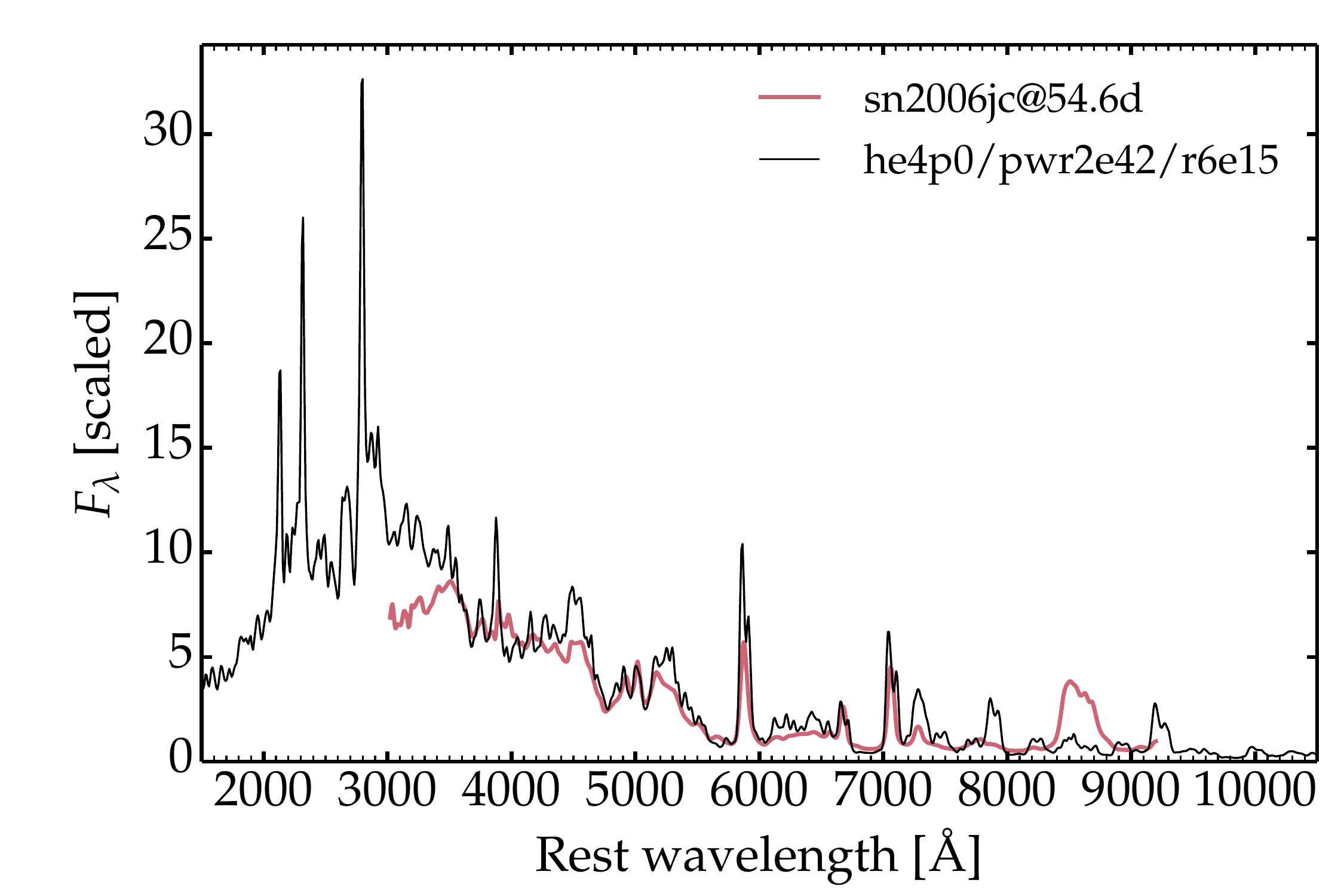}
\caption{Left: Comparison between the observation of SN\,2006jc on the 14th of October 2006, 14\,d after the estimated time of maximum at $MJD=$\,54008, with the spectrum from a CDS whose mass and composition are based on model he4p0, placed at a radius of $2 \times 10^{15}$\,cm, moving at a velocity of 2000\,\kms\ and powered at a rate of $6 \times 10^{42}$\,\ergs. Right: Same as left but now for the observations of SN\,2006jc on 23rd of November 2006, 54.6\,d after the estimated time of maximum, and for the same he4p0 model in a CDS at $6 \times 10^{15}$\,cm, moving at a velocity of 2000\,\kms\ and powered at a rate of $2 \times 10^{42}$\,\ergs. In both panels, the observations have been corrected for reddening ($E(B-V)=$\,0.04\,mag) and redshift ($z=$\,0.005574). The model is normalized to the observations at 5300\,\AA\ (left panel) and at 5800\,\AA\ (right panel). Models and observations are smoothed with a Gaussian kernel (FWHM of 23.5\,\AA).
\label{fig_06jc}
}
\end{figure*}
\begin{figure}
\centering
\includegraphics[width=\hsize]{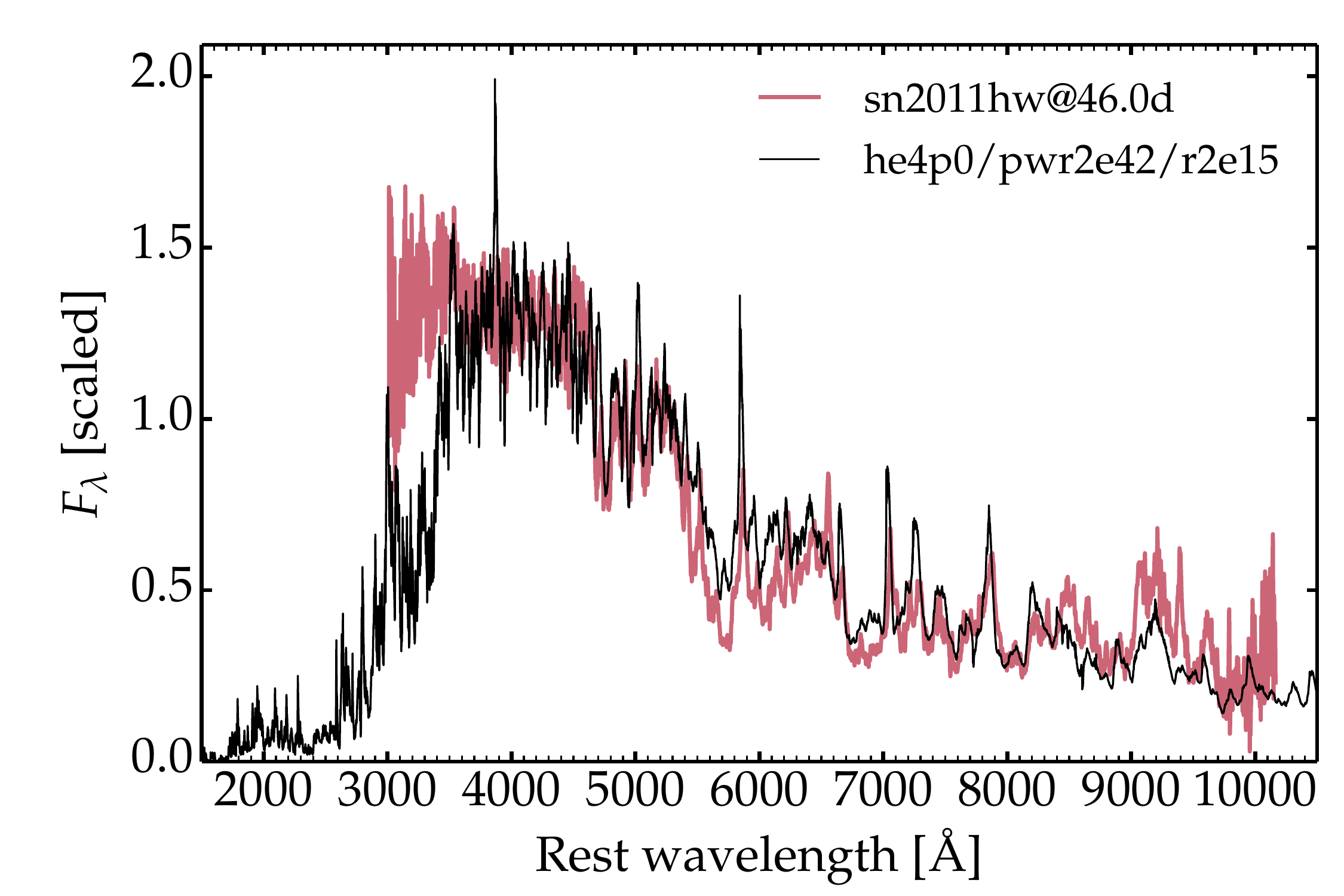}
\caption{Comparison between the observation of SN\,2011hw on the 21st of December 2011, 46\,d after the estimated time of maximum at $MJD=$\,55870, with the spectrum from a CDS whose mass and composition are based on model he4p0, placed at a radius of $2 \times 10^{15}$\,cm, moving at a velocity of 2000\,\kms\ and powered at a rate of $2 \times 10^{42}$\,\ergs. The observed spectrum has been corrected for reddening ($E(B-V)=$\,0.115\,mag) and redshift ($z=$\,0.023) and the model is normalized to the observations at 5300\,\AA. No smoothing was applied either to the observation or the model. The data and characteristics of SN\,2011hw are taken from \citet{pasto_ibn_trans_15}.
\label{fig_11hw}
}
\end{figure}

\section{Comparison to observations}
\label{sect_obs}

We now compare the models discussed in the previous sections with the observations of a few Type Ibn SNe. No model was tailored to any specific Type Ibn SN. The choice is vast and the scenario explored here may not apply to all observed Type Ibn SNe. In particular, one critical feature predicted in our models (excluding those with a nearly pure helium composition) is strong Fe\two\ emission below 5500\,\AA\ so Type Ibn SNe with no such emission cannot be satisfactorily matched by our models. Our models seem to capture some essential features of the prototypical Type Ibn SN  2006jc \citep{pasto_06jc_07}, as well as the Type Ibn SNe with relatively weak narrow lines like  2011hw \citep{pasto_ibn_trans_15} and 2018bcc \citep{karamehmetoglu_18bcc_21}. The comparisons in this section are limited to these events, while other Type Ibn SNe will be studied in forthcoming studies.

As we emphasize in Sect.~\ref{sect_rt}, different model parameters can yield similar spectra. For example, a higher CDS mass at a larger  radius may yield a similar spectrum to that from a lower CDS mass at a smaller radius. A spectrum may remain unchanged if the shock power is reduced as the CDS radius is increased etc. In general, a combination of effects are at work and spectral properties tend to be similar when the ionization is the same. There are thus many degeneracies. The goal here was to identify what CDS properties were necessary to produce a spectrum that resembled Type Ibn SNe like 2006jc and analogs. Developing a global model for such events (i.e., matching the multiband light curves and multiepoch spectra) is left to future work.

We correct the spectra for reddening and redshift, as described in those papers. The power adopted in our models is not adjusted to match the inferred SN flux, although it does not differ by much, so we normalize our model spectra to the observed flux, typically at 5500\,\AA. In general, the model spectra are smoothed with a Gaussian Kernel (FWHM of 23.5\,\AA) in order to reduce the offset with the spectral resolution of the observations, to mimic the potentially large turbulence within the CDS (our simulations adopt a turbulent velocity of 50\,\kms), and to smoothen the sometime noisy observed spectra.

\subsection{SN\,2006jc}
\label{sect_06jc}

Figure~\ref{fig_06jc} presents a comparison of our CDS models, based on the composition and mass of model he4p0, with the observations of SN\,2006jc at 14.0\,d (left panel) and 54.6\,d (right panel) after the time of maximum.
In the context of interaction, the shock decelerates as material piles up in the CDS (see, for example, the illustration in Fig.~\ref{fig_rhd} for model E5). So, as the shock power drops in time (the more so for a stronger deceleration), the CDS mass grows, and the CDS velocity drops (see also discussion for a large set of simulations for Type IIn SNe in \citealt{D15_2n,D16_2n}). Because the CDS moves, the CDS radius increases in time. So, when studying the properties of Type Ibn SNe are different epochs, it is natural to invoke different powers and CDS properties. The values used here for SN\,2006jc may not be strictly correct since they are taken from the grid of models presented in Sect.~\ref{sect_rt}, which were performed with no specific SN in mind.

A greater power and a smaller radius is used for the first epoch, leading to a bluer optical spectrum, a stronger continuum flux, and weaker lines. The model reproduces the overall shape and a number of observed spectral signatures are also predicted by the model. The overall shape of the spectrum is due to Fe\,\two\ emission already at that time, although most of the Fe\two\ emission occurs below 4000\,\AA\ -- we are mostly seeing the tail of the emission in the optical.  The main He\one\ lines predicted by the model and in general also observed are at 3889, 3965, 4026,  4471, 5016, 5876, 6678, 7065, and 7281\,\AA. The He\one\,3889\,\AA\ line is not observed while the model predicts it. This may be due to the blanketing effect of Fe\two\ on this He\one\ line, which is weak in the model because He\one\,3889\,\AA\ is predicted to form in a large part in the outer regions of the emitting region. In reality, chemical mixing may cause a more intertwined emission of He\one\ and Fe\two\ lines, causing a greater blanketing of He\one\ lines by Fe\two\ lines.

The strongest predicted  O\,\one\ lines are multiplets around 8446\,\AA, and there are also O\,\two\ multiplets around 7320\,\AA. Two sets of strong Si\two\ lines are present at 5041--5056, 5958--5979, and 6347-6371\,\AA. Strong Mg\two\, lines are predicted at 4481, 7896, 8234, 9218, and 9244\,\AA, but these are observed with a much lower strength. Overall, the model underestimates the width of line profiles (this does not seem to be a resolution issue since the observed spectral lines in this early-time spectrum are indeed broader than at later times, with the exception of the Ca\two\ near-infrared triplet),  probably because the spectrum forms over a range of velocities and densities rather than in a narrow CDS as assumed. This assumption holds best at late times when ejecta and CSM have all been swept up into a dense confined region.

The right panel of Fig.~\ref{fig_06jc} shows a comparison of SN\,2006jc at 54.6\,d after maximum with the CDS now at a larger radius of 6 rather than $2 \times 10^{15}$\,cm and powered at a rate of 2 rather than $6 \times 10^{42}$\,\ergs. These parameters are those used in the model discussed in Sect.~\ref{sect_he4p0}. The model compares satisfactorily in terms of Fe\two\ emission in the blue, the presence of numerous He\one\ lines, but the model overestimates the strength of Mg\two\ lines and underestimates the strength of Ca\two\ lines. This discrepancy may arise from an inadequate density structure (for example due the presence of clumping, which is neglected here). The CDS is also expected to have a very complicated structure, probably with multiple emission regions having a range of density compressions, ionization, temperature, turbulent velocity etc. Obtaining a better model will require more work, and in particular a better representation of the CDS and the interaction region based on realistic 3D radiation-hydrodynamics simulations (which are currently lacking).

\subsection{SN\,2011hw}
\label{sect_11hw}

Figure~\ref{fig_11hw} compares the observations of SN\,2011hw at 46\,d after the inferred time of explosion with the model for a CDS composition and mass given by model he4p0, a radius of $2 \times 10^{15}$\,cm, a velocity of 2000\,\kms, and powered at a rate of $2 \times 10^{42}$\,\ergs. The observations are corrected for reddening and for redshift adopting $E(B-V)=$\,0.115\,mag and $z=$\,0.023; the adopted time of explosion is $MJD=$\,55870 \citep{pasto_ibn_trans_15}.

The agreement is  better than for SN\,2006jc, in particular for metal lines such as those of Mg\two\ or Ca\two. Compared to the  54.6\,d spectrum of SN\,2006jc, the 46.0\,d spectrum of SN\,2011hw shows stronger Mg\two\ lines and weaker Ca\two\ lines. The observed flux in the blue is well matched by the Fe\two\ emission in the model.

\begin{figure}
\centering
\includegraphics[width=\hsize]{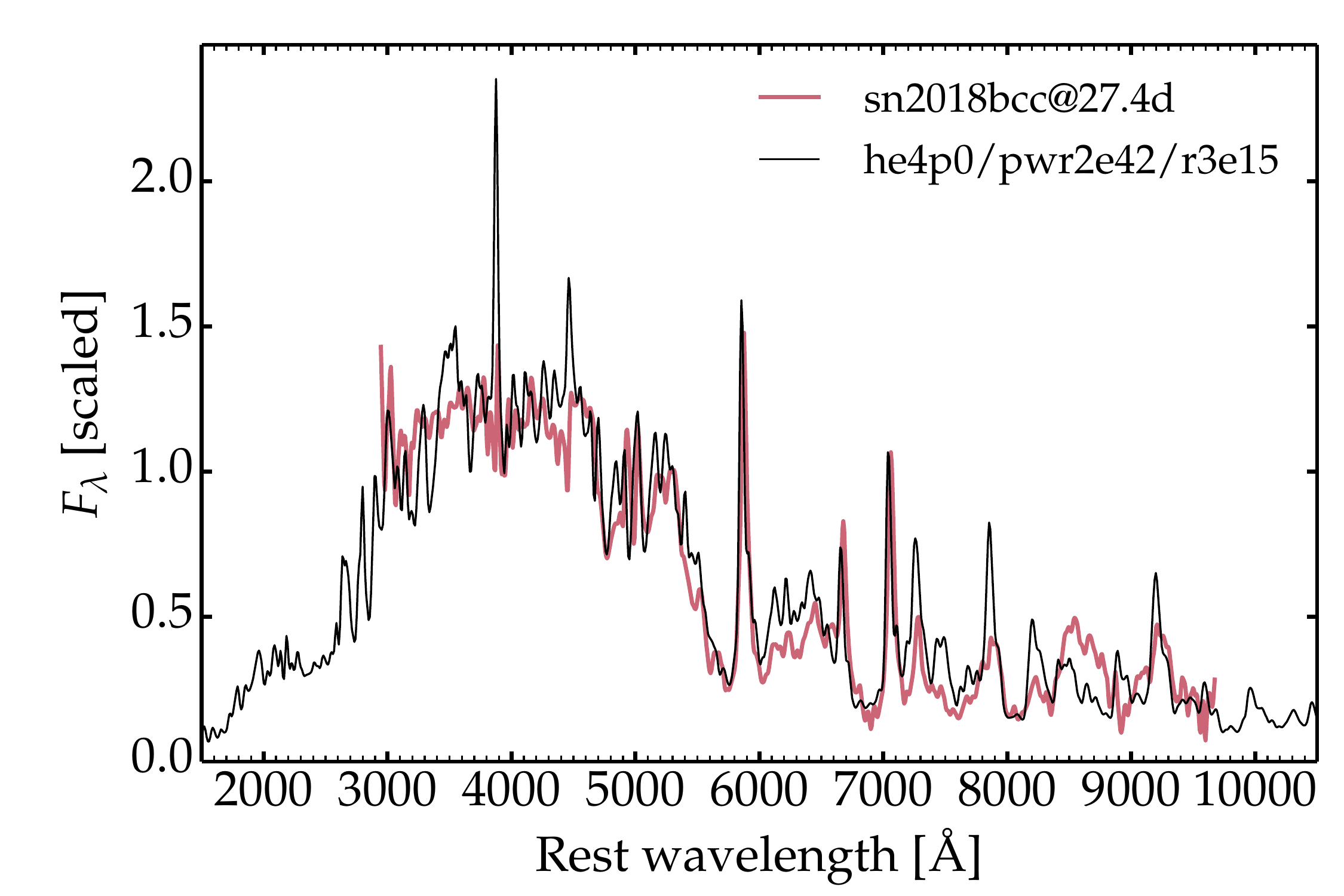}
\caption{Comparison between the observation of SN\,2018bcc on the 14th of May 2018, 27.4\,d after the estimated time of maximum at $MJD=$\,58225.5, with the model spectrum based on a CDS mass and composition based on model he4p0, placed at a radius of $3 \times 10^{15}$\,cm, moving at a velocity of 2000\,\kms\ and powered at a rate of $2 \times 10^{42}$\,\ergs. The observed spectrum has been corrected for reddening ($E(B-V)=$\,0.0124\,mag) and redshift ($z=$\,0.0636) and the model is normalized to the observations at 5300\,\AA. Both are smoothed with a Gaussian kernel (FWHM of 23.5\,\AA). The data and characteristics of SN\,2018bcc are taken from \citet{karamehmetoglu_18bcc_21}.
\label{fig_18bcc}
}
\end{figure}

\subsection{SN\,2018bcc}
\label{sect_18bcc}

Figure~\ref{fig_18bcc} compares the observations of SN\,2018bcc on the 14th of May 2018, 27.4\,d after the estimated time of explosion at $MJD=$\,58225.5 \citep{karamehmetoglu_18bcc_21}, with the model spectrum that adopts a CDS mass and composition based on model he4p0, placed at a radius of $3 \times 10^{15}$\,cm, moving at a velocity of 2000\,\kms\ and powered at a rate of $2 \times 10^{42}$\,\ergs. The comparison is a close analog to that shown for SN\,2011hw but serves to show that despite the diversity of Type Ibn SNe, there is also a strong similarity in spectral properties, in particular a few weeks after the light curve maximum.

Although the features are not always well matched in strength, most of the observed features are predicted in the model. Just like for SNe 2006jc and 2011hw, the Mg\two\ lines are overestimated and the Ca\two\ lines are too weak, suggesting an ionization offset in the spectrum formation region.

\section{Conclusion}
\label{sect_conc}

We have presented numerical simulations for Type Ibn SNe under the assumption that these events are the result of the interaction between the ejecta of an exploding star with the CSM produced in a pre-SN outburst either in the form of a dense wind or a non-terminal explosion unbinding only part of the star. Radiation-hydrodynamics simulations for such configurations suggest that a wide variety of light curves and CDS properties may be produced, spanning a large range of peak luminosities up to the most luminous SNe known. However, we find that in order to produce a peak luminosity and narrow spectral lines compatible with the observations of numerous Type Ibn SNe like 2006jc, one requires a low-energy low-mass ejecta ramming into a massive slow-moving outer shell (a similar conclusion is proposed by \citet{moriya_ibn_16} based on the study of Type Ibn SN bolometric light curves). This contrasts with the general consensus on Type Ibn SNe, which invokes an energetic Wolf-Rayet star explosion embedded in a dense CSM. Such a configuration conflicts with the modest peak luminosity and the absence of broad lines at all epochs in events like SN\,2006jc.

A physically consistent model for a Type Ibn SN like 2006jc could be a helium star in the initial mass range between 2.6 and 3.3\,\msun\ \citep{ertl_ibc_20}. Such progenitors have attractive properties. Their envelope is significantly expanded to 10$^{12}$--10$^{13}$\,cm at the end of the star life so that the outer 1\,\msun\ is weakly bound. They burn oxygen and silicon via convectively bounded flames that might potentially be more unstable. Such progenitors can have silicon flashes and eject an unknown amount of mass. Following the flash, the remaining star explodes and produces a low-mass ejecta, potentially endowed with a few 0.01\,\msun\ of \nifs\ (Stan Woosley, private communication).

The interaction model that seems most suitable for events like 2006jc eventually develops into a very simple structure where the bulk of the swept-up CSM and the decelerated ejecta material has piled up into a CDS. Such a configuration is amenable to a relatively local radiative transfer calculation that assumes that the bulk of the SN radiation arises from the CDS under the influence of some power (i.e., a mixture of shock interaction and radioactive decay). In the absence of a good model for the Type Ibn SN scenario, we further assumed that such a CDS has a composition and total mass similar to the helium-star explosions of \citet{ertl_ibc_20}. This approach is flexible and allows for broad investigations into the nature of Type Ibn SN progenitors and interaction properties.

Non-LTE radiative transfer calculations indicate that the late time spectra of Type Ibn SNe like 2006jc, 2011hw, or 2018bcc are dominated by Fe\two\ emission below 5500\,\AA, with strong isolated lines of He\one\ if the fractional helium abundance is above about 50\%.  Our models tend to overestimate the ionization level of metals in the CDS, with Mg\two\ lines stronger than observed but Ca\two\ lines weaker than observed. A grid of simulations based on 2.9 to 12\,\msun\ helium-star models  reveal the continuous weakening of He\one\ lines, the progressive reddening of the spectra, and the ever strengthening Fe\two\ emission. The best agreement to Type Ibn SNe like  2006jc, 2011hw, or 2018bcc is obtained for a CDS composition and mass representative of  a relatively low-mass helium-star explosion model (here he4p0), with about 1\,\msun\ of helium and the rest being distributed amongst carbon, oxygen, neon, magnesium, and heavier metals.

The iron content of the CDS has a critical impact. We found that the iron from \nifs\ decay plays a role, but more important is the metallicity of the progenitor star since that impacts all metals rather than just Co and Fe. At one tenth solar, the metal emission below 5500\,\AA\ disappears, producing a blue spectrum with strong lines of He\one\ in the he4p0 model. Since this Fe\two\ emission is a recurrent feature of Type Ibn SNe, it suggests that observed Type Ibn SNe arise from solar-metallicity stars. This seems to contrast with expectations for pulsational-pair instability in very massive stars, despite their promising properties. Hence, currently observed Type Ibn SNe compare favorably with the expectations for solar-metallicity lower-mass helium stars in binary systems and undergoing non-terminal eruptions a short while before core collapse.

All simulations presented here assume spherical symmetry. Departures from spherical symmetry may exist on both small and large scales and affect both the inner shell and outer shells. With the assumption of spherical symmetry, the parameter space is already vast. Allowing for asymmetry increases this parameter space  dramatically. On small scales, the dense shell should break up and produce a complex, clumpy structure (see, for example, the exploratory simulations by \citealt{blondin_sn2n_96}). This could lead to the simultaneous present of material with different temperatures, ionization, and composition within the CDS. In our study, we have assumed some chemical  mixing  but neglected clumping. Detailed 3D radiation-hydrodynamics simulations are necessary to explore this aspect. Asymmetries on large scales have already been studied, for example with the same \heracles\ code by \citet{vlasis_2n_16}. This does not lead to any dramatic differences unless the asymmetry is extreme. In the context of interactions, one should recall that the engine is only efficient if the CSM completely enshrouds the incoming ejecta. If interaction occurs only over a restricted solid angle, the interacting power and consequently the SN luminosity is reduced. Furthermore, one may obtain a hybrid spectrum with both narrow and broad lines. There does not seem to be strong evidence for large asymmetry in Type Ibn SNe so exploring spherical models remains a sensible first step.

 The present work should not mislead the reader in thinking that all Type Ibn SNe arise from the helium-star progenitors studied here. The Type Ibn SN diversity is so vast that multiple formation channels are likely. With the composition of helium-star progenitors, our model spectra tend to overestimate (underestimate) the strength of Mg\two\ (Ca\two) lines compared to what is observed, for example, in SN\,2006jc. While we cannot exclude that the discrepancy arises from an inadequate density structure in the CDS (for example, because of clumping, which we neglect), this discrepancy may indicate a composition offset. A configuration more rich in calcium could perhaps resolve this issue. Helium detonations at the surface of white dwarfs produce a helium-rich and calcium-rich composition \citep{waldman_hedet_11} that may be more suitable for some Type Ibn SNe, in particular for events located in environments that lack any obvious star  formation \citep{hosseinzadeh_ibn_19}. A comprehensive and systematic modeling of the full diversity of Type Ibn SNe is needed.

\begin{acknowledgements}

We thank Stan Woosley, Raffaella Margutti and Andrea Pastorello for useful discussions. This work was supported by the ``Programme National de Physique Stellaire'' of CNRS/INSU co-funded by CEA and CNES. DJH thanks NASA for partial support through the astrophysical theory grant 80NSSC20K0524. H.K. was funded by the Academy of Finland projects 324504 and 328898. This work was granted access to the HPC resources of  CINES under the allocation  2019 -- A0070410554 and 2020 -- A0090410554 made by GENCI, France. This research has made use of NASA's Astrophysics Data System Bibliographic Services.

\end{acknowledgements}

\begin{appendix}

\section{Additional information on the radiation-hydrodynamics simulations with \heracles}

\begin{figure}
\centering
\includegraphics[width=\hsize]{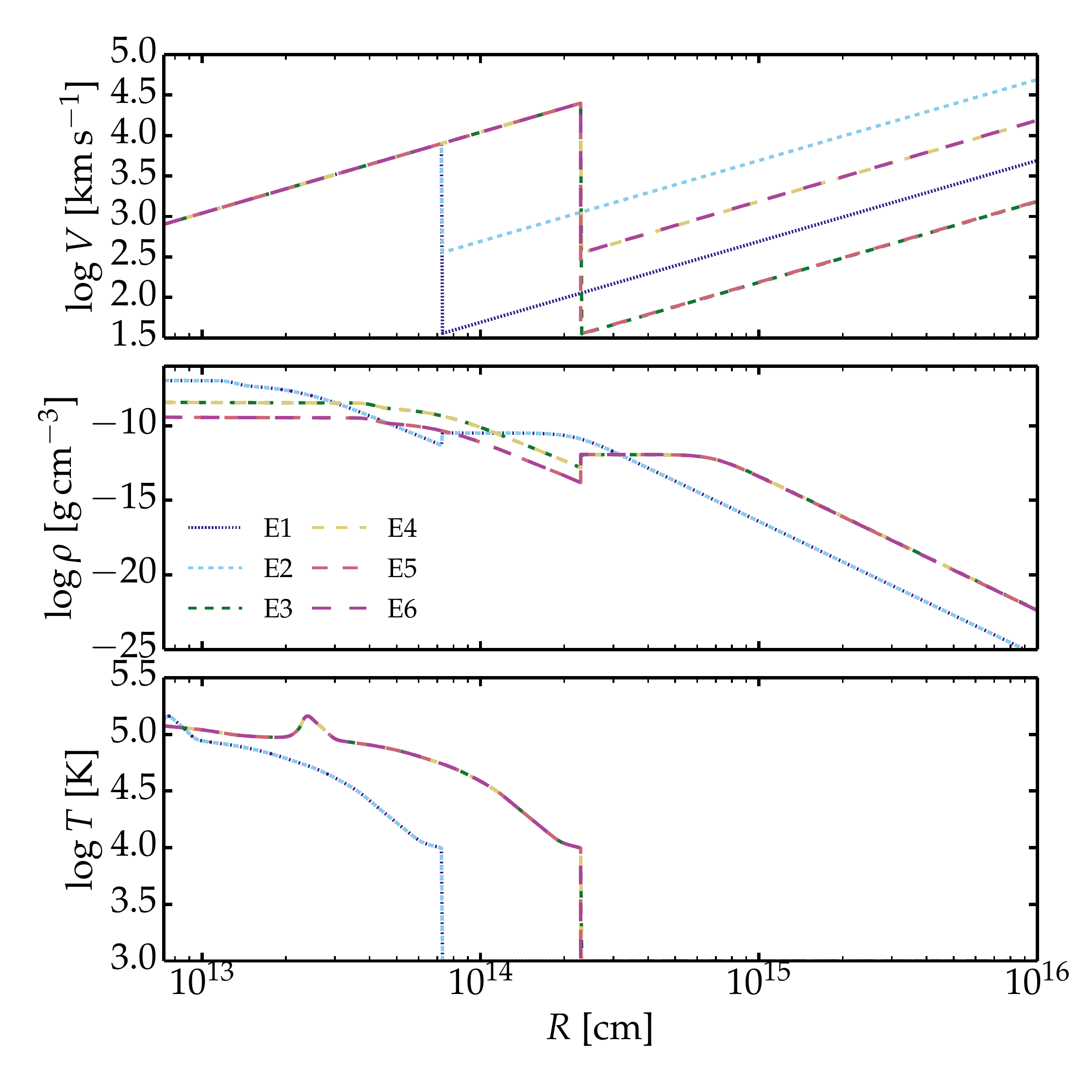}
\includegraphics[width=\hsize]{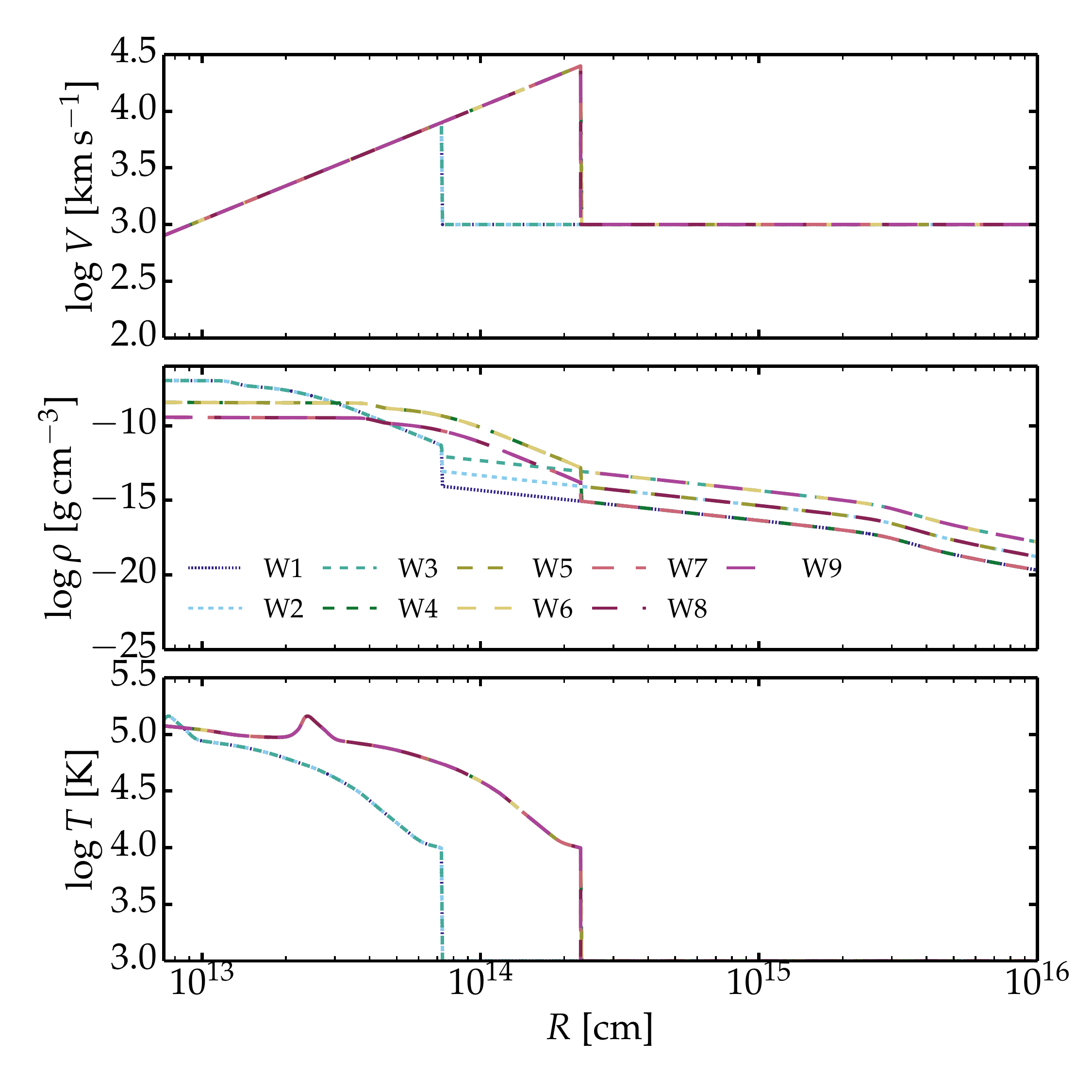}
\vspace{-0.5cm}
\caption{Illustration of the initial radial profile of the velocity, density, and temperature for all interaction configurations studied in Sect.~\ref{sect_rhd} and described in Table~\ref{tab_her}. Three types of inner shell properties are considered, with the unadulterated model he4 and two scaled variants (in density, and in both density and velocity). For the outer shell, we consider an ejecta CSM with two different kinetic energies or a wind CSM with three different mass loss rates.
\label{fig_prop_init}
}
\end{figure}

Figure~\ref{fig_prop_init} illustrates the various configurations for which we performed multi-group radiation-hydrodynamics simulations with \heracles. The junction between inner shell and outer shell is easily seen in each panel as a sharp jump in velocity, density or temperature.

\begin{figure*}
\centering
\includegraphics[width=0.495\hsize]{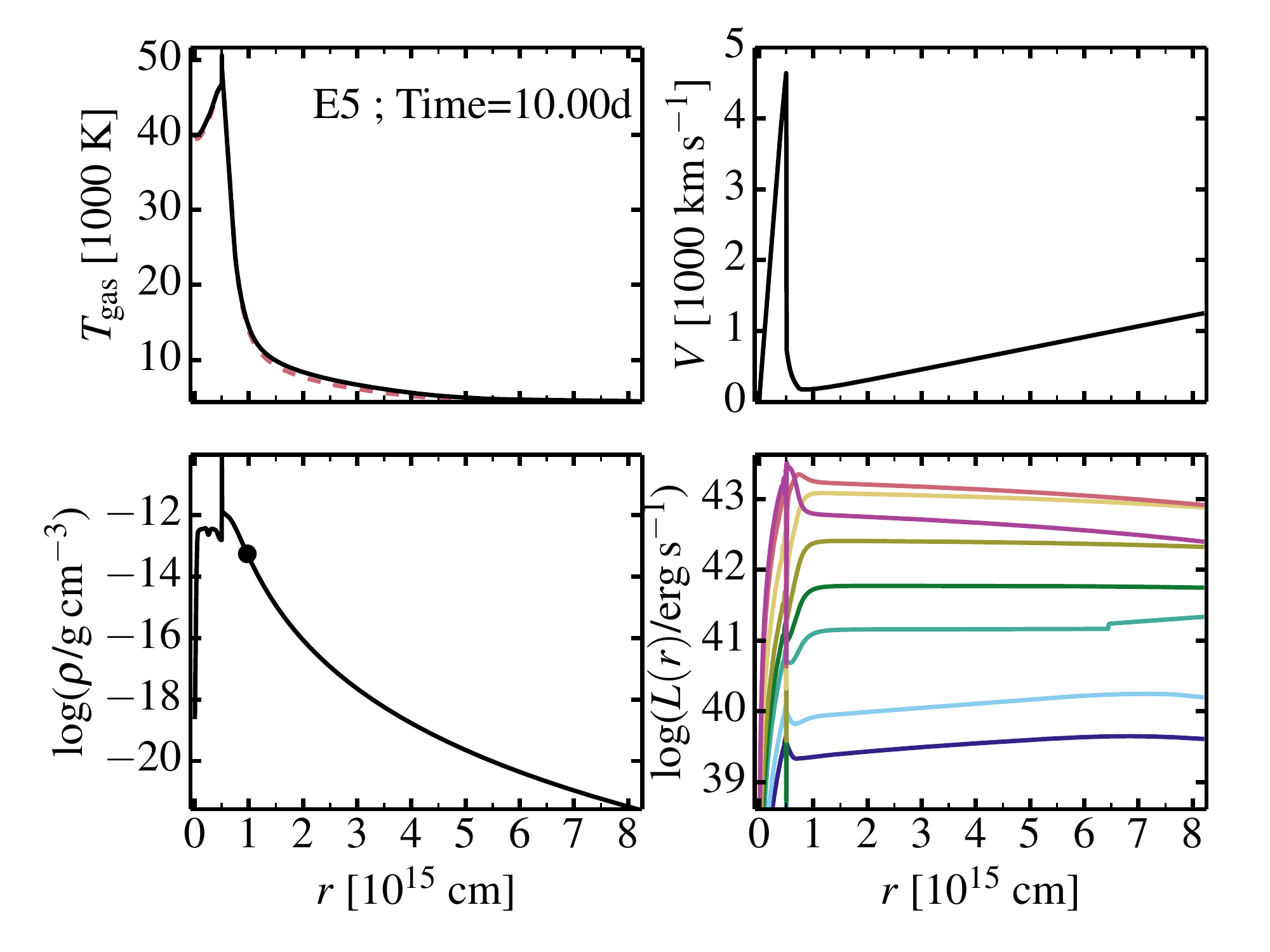}
\includegraphics[width=0.495\hsize]{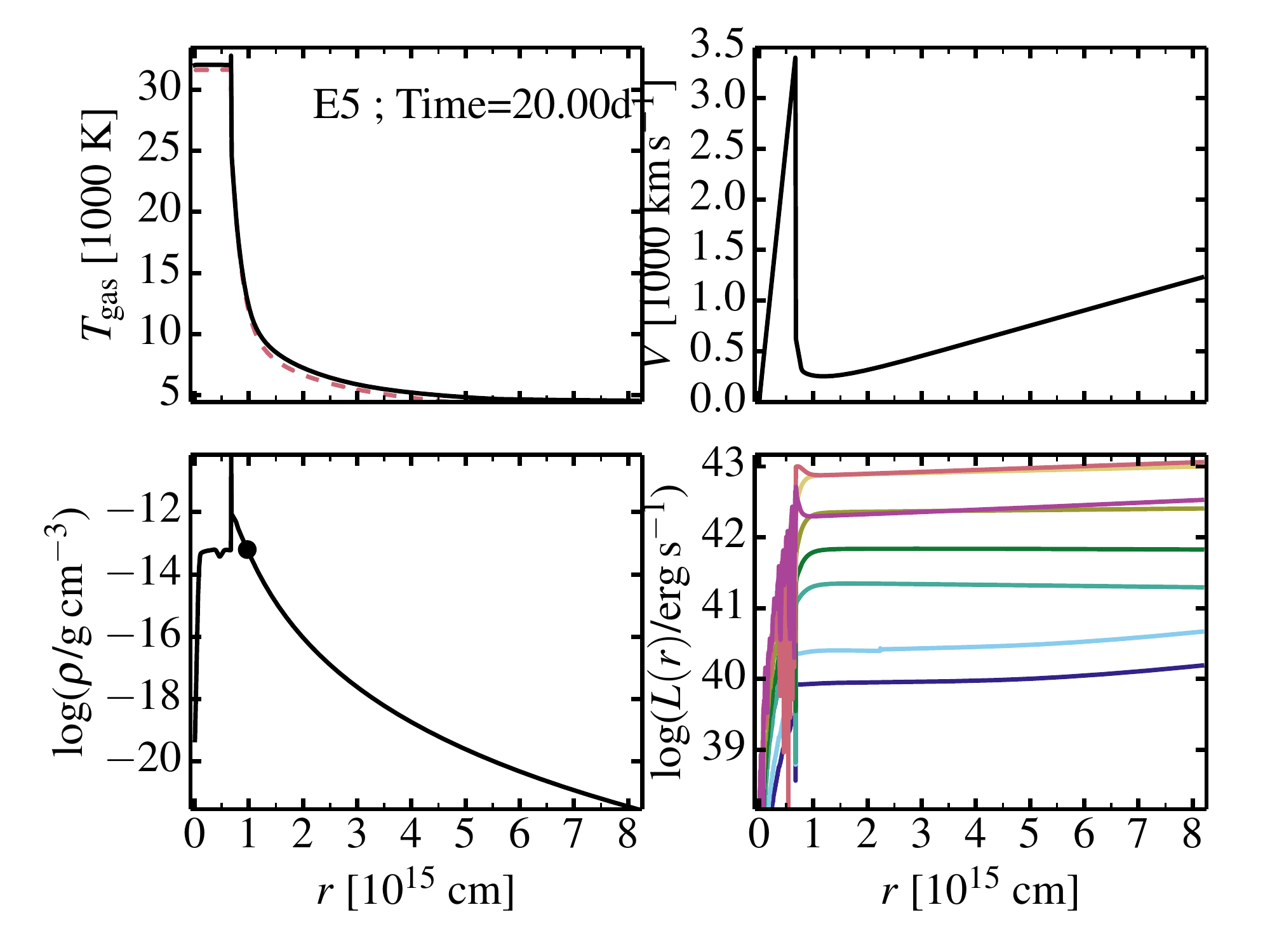}
\includegraphics[width=0.495\hsize]{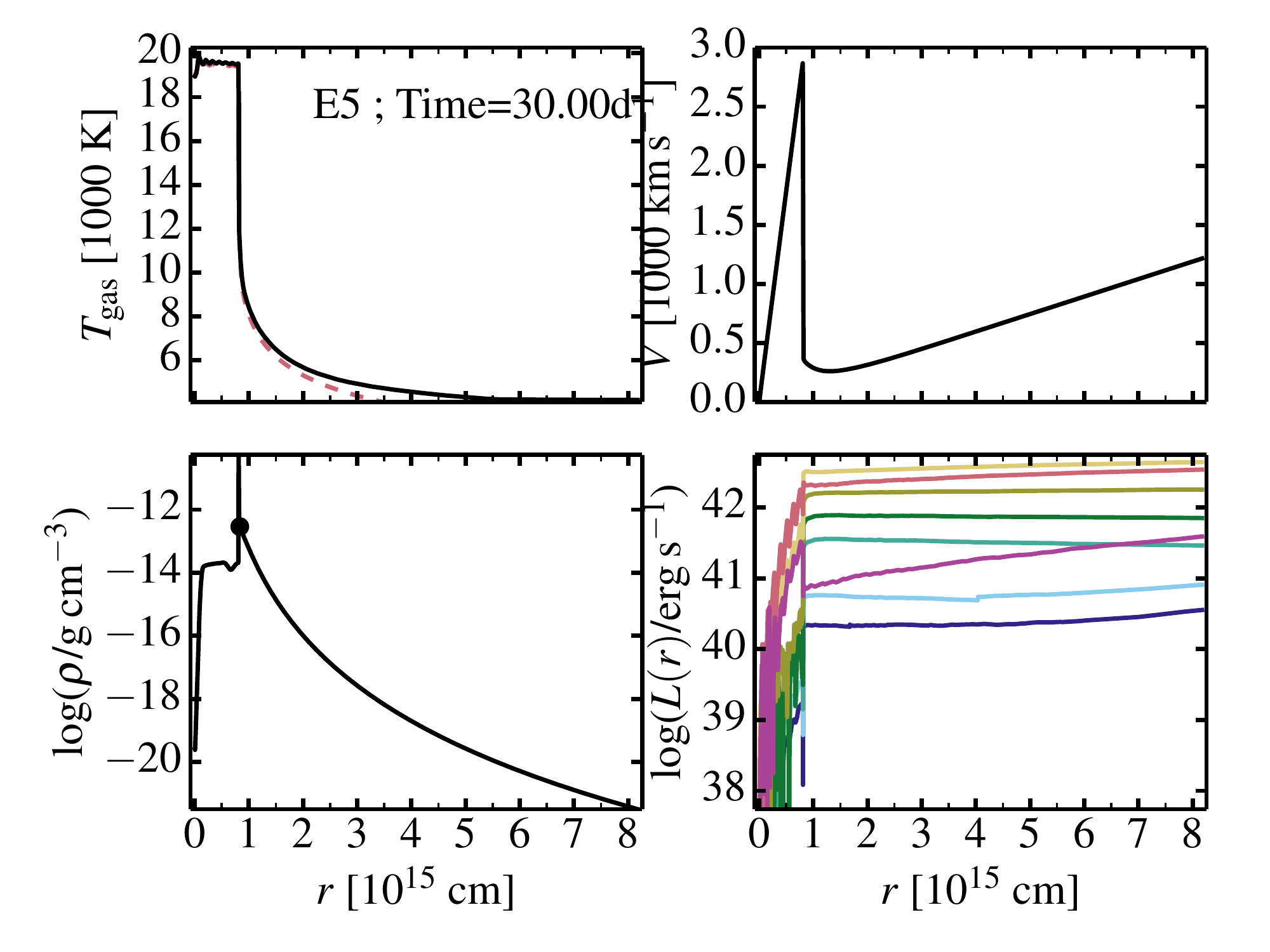}
\includegraphics[width=0.495\hsize]{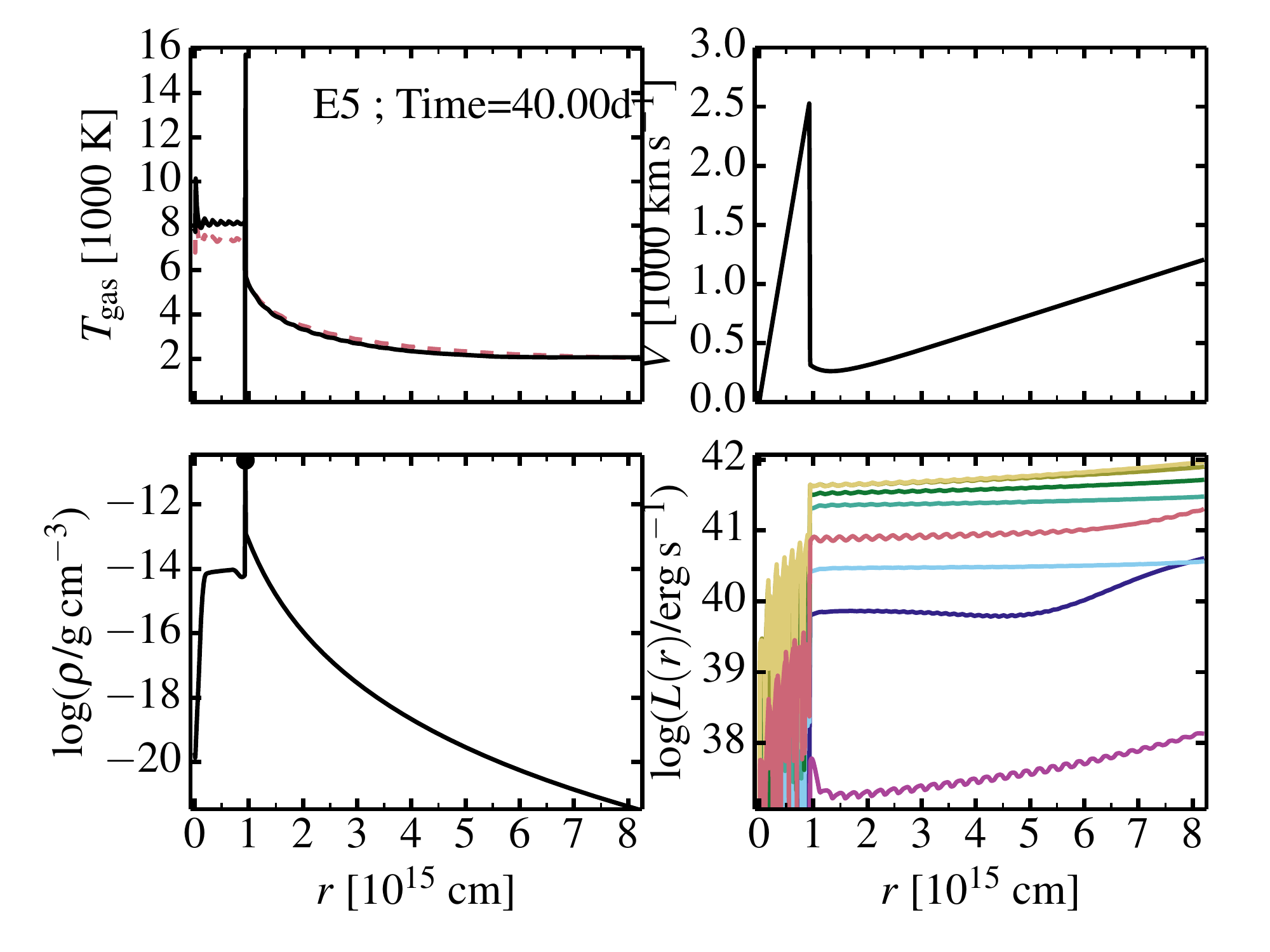}
\includegraphics[width=0.495\hsize]{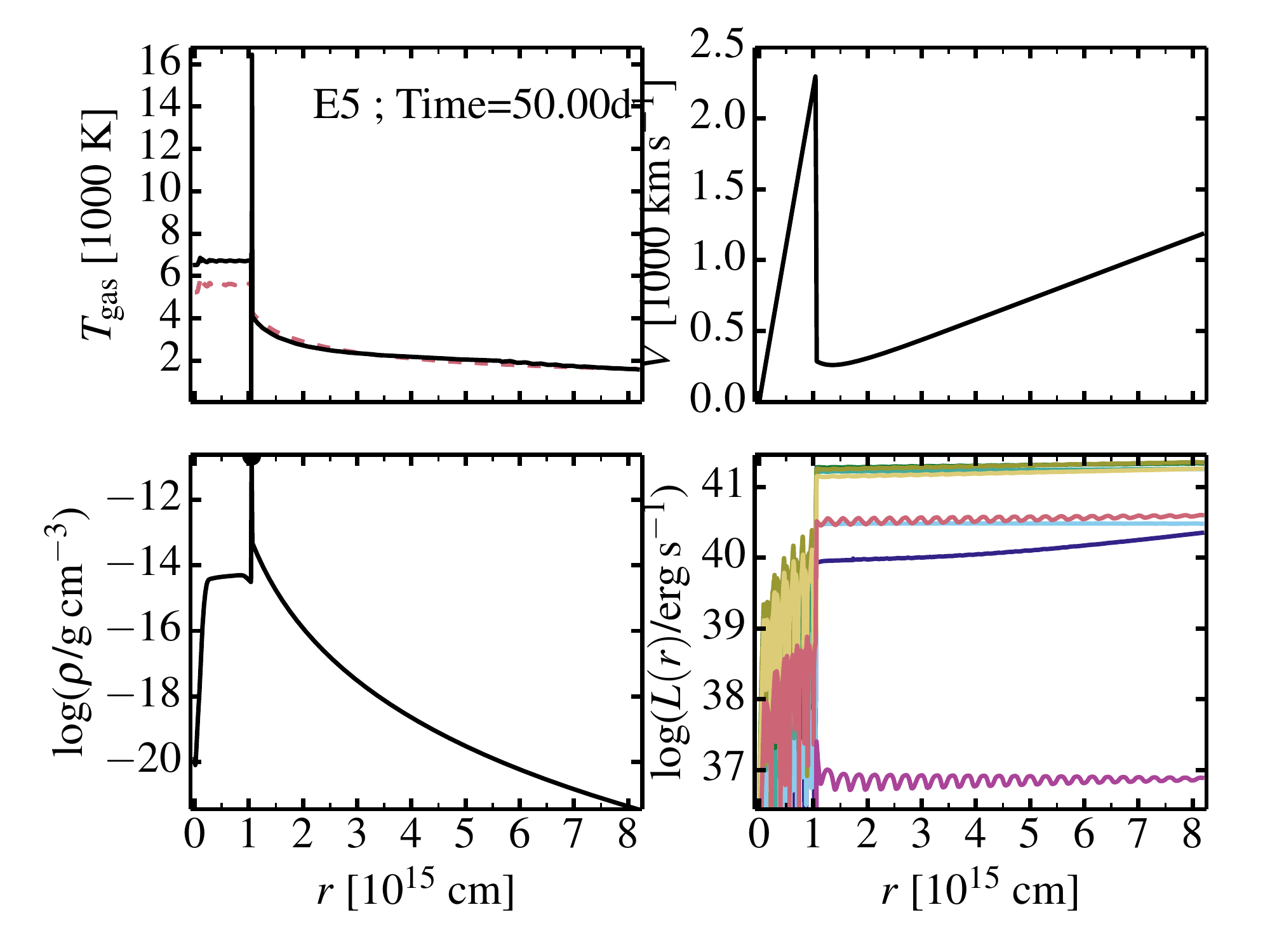}
\vspace{-0.2cm}
\caption{Radial profile of the gas temperature $T_{\rm gas}$, the velocity $V$, the mass density $\rho$, the luminosity $L(r)$ for model E5 at 10, 20, 30, 40, and 50\,d after the onset of interaction. The black dot in the bottom left panel indicates the location of the electron-scattering photosphere (where the inward integrated electron-scattering optical depth is 2/3).
\label{fig_rhd}
}
\end{figure*}

Figures~\ref{fig_rhd} illustrates the evolution of the interaction for model E5, whose initial configuration and bolometric light curve are shown in Fig.~\ref{fig_prop_init} and Fig.~\ref{fig_lbol_her}, respectively. A velocity jump separates the inner and outer shells at all times, i.e. some shock power is injected in this configuration at all times. Because the density in the standard model he4p0 from \citet{dessart_snibc_20} was scaled by 0.1, there is only 0.008\,\msun\ so the decay power is a small power source at all times. The mass contrast between inner and outer shell (0.15 versus 1\,\msun), there is strong deceleration of the inner shell material, which leads to a substantial conversion of kinetic energy (initial budget was $7.5 \times 10^{49}$\,erg) into radiative energy (the time-integrated bolometric luminosity is $5.7 \times 10^{49}$\,erg). The peak luminosity reached by model E5 at 11.4\,d is $4.2 \times 10^{42}$\,\ergs.

In this model, the unshocked CSM stays optically thick for about 20\,d. This probably means the emergent radiation would be reprocessed entirely by the unshocked CSM for about 10\,d. After that, the radiation would form from both the unshocked CSM and the CDS, until a time when it would entirely form in the CDS. Indeed, at 50\,d after the onset of the interaction, the photosphere is within the CDS and most of the inner-shell and outer-shell material has piled up inside the CDS, which is moving at about 1400\,\kms. These properties are  compatible with those inferred from the Type Ibn SNe 2006jc, 2011hw or 2018bcc.

\end{appendix}

\end{document}